\documentclass[%
 prfluids,
 reprint,
 onecolumn,
 superscriptaddress,
 aps,
 nofootinbib
]{revtex4-2}
\usepackage{silence}
\usepackage[T1]{fontenc}
\usepackage{graphicx}
\usepackage{dcolumn}
\usepackage{bm}
\usepackage{float}
\usepackage{amsmath,amssymb,amsfonts}
\usepackage{todonotes}
\usepackage{epstopdf}
\usepackage{multirow}
\usepackage{longtable}

\usepackage{comment}
\usepackage{cancel}
\usepackage[normalem]{ulem}

\def\XXint#1#2#3{{\setbox0=\hbox{$#1{#2#3}{\int}$}
\vcenter{\hbox{$#2#3$}}\kern-.5\wd0}}

\def\XXiint#1#2#3{{\setbox0=\hbox{$#1{#2{\mathrm{#3\!\! #3}}}{\iint}$}
\vcenter{\hbox{$#2 {\mathrm{#3\!\! #3}}$}}\kern-.5\wd0}}
\usepackage{enumitem}
\usepackage{hyperref}
\usepackage{color}  
\definecolor{darkGreen}{rgb}{0,0.45,0}
\definecolor{darkBlue}{rgb}{0,0,0.7}
\definecolor{darkRed}{rgb}{0.76, 0.13, 0.28}
\hypersetup{
	colorlinks=true, 
	linktoc=all,    
	linkcolor=darkBlue, 
	citecolor=darkGreen,
	urlcolor = darkRed,
}
\usepackage{comment}

\newcommand{\upd}{\mathrm{d}}

\usepackage{overpic}

\renewcommand{\d}{{\mathrm{d}}}

\usepackage[margin=2.3cm]{geometry}
\begin{abstract}
Long arrays of identical, self-propelling flapping flyers are inherently unstable and thus unlikely to exist without active control mechanisms. One approach to enable long in-line formations is to enforce a constant separation between the group members. The objective then becomes to determine the flapping strategies the flyers should adopt to achieve a certain separation. Using an aerodynamic model of vortex wake production and inter-flyer effects, we explore different flapping strategies for followers given the motion of the leader. The choice of tactic is dependent upon the aerodynamic, kinematic, and physical parameters of the system, and reflects an interplay between efficiency and stability. We find that whether a flyer flaps in or out of phase with its upstream neighbour, together with the target separation, strongly affect the flapping amplitude and, therefore, the energetic cost of group flight. In certain regimes, group flight is energetically favourable compared to isolated flight, while in others, flying in formation becomes less efficient. We also identify ``goldilocks zones'', ranges of separation in which one of the in- or out-of-phase motions can be simultaneously energetically efficient and dynamically stable. Outside these regions, energetically favourable flight is unstable and therefore unlikely to occur.
\end{abstract}

\begin{document}

\title{Flapping strategies for flying formations}
\author{Javier Chico-V\'azquez}
\author{Christiana Mavroyiakoumou}
\email[Electronic address: ]{christiana.mavroyiakoumou@maths.ox.ac.uk}
\affiliation{Mathematical Institute, University of Oxford, Woodstock Road, Oxford, OX2 6GG, UK}

\maketitle

\section{Introduction}

Flow-mediated interactions play an important role in the emergent structure of flying and swimming formations and are a main ingredient to consider when studying collective locomotion~\cite{sumpter2006principles,cavagna2014bird,hemelrijk2012schools}.
The first to propose that such flow-mediated forces of interaction among individuals in a group could be responsible for the structure of animal collectives was Sir James Lighthill~\cite{lighthill1975mathematical}. His conjecture has drawn a large amount of interest among researchers who try to characterise such systems~\cite{zhu2014flow,becker2015hydrodynamic,ramananarivo2016flow,dai2018stable,peng2018hydrodynamic,newbolt2019flow,newbolt2024flow}.

In this work, we focus on in-line formations in which flyers are arranged one after another and whose motions along this line are determined dynamically through their communication via the surrounding flow environment. This idealised arrangement provides a simple setting for studying wake interactions among bodies that self-propel and their effects on the group dynamics~\cite{hemelrijk2012schools,becker2015hydrodynamic,peng2018hydrodynamic,oza2019lattices,saadat2021hydrodynamic,heydari2021school,newbolt2024flow,nitsche2025stability}. The interactions in this setup occur through the flows generated by a flyer that oscillates in the vertical direction, and is considered representative of high Reynolds number flapping locomotion~\cite{triantafyllou1993optimal,andersen2017wake,becker2015hydrodynamic,ramananarivo2016flow}. Flapping flight in linear formations also allows for the design of experiments and the development of fluid dynamical simulations on actuated foils or wings~\cite{becker2015hydrodynamic,ramananarivo2016flow,newbolt2019flow,newbolt2024flow}.
Some previous studies considered the thrust generation and efficiency of a single flapping foil~\cite{triantafyllou1993optimal} or a pair of flapping foils in tandem~\cite{boschitsch2014propulsive,rival2011recovery} fixed within an oncoming flow, while other works focused on the free locomotion of flapping foils in forward flight~\cite{vandenberghe2004symmetry,alben2005coherent,vandenberghe2006unidirectional,spagnolie2010surprising,ramananarivo2016flow,newbolt2019flow,newbolt2024flow,han2025tailoring}.

In the majority of these studies, the flyers undergo a prescribed vertical motion, without sensing or any ability to adapt their kinematics. In most cases, the flyers also flap in a coordinated manner, following identical amplitudes and frequencies~\cite{ramananarivo2016flow}. Systems of uncoordinated flyer pairs have been examined as well~\cite{newbolt2019flow}. In that case, flapping with differing kinematics can still spontaneously lead to group cohesion, with differing temporal phases yielding equilibrium states of different separations~\cite{newbolt2019flow,mavroyiakoumou2025modeling}. The stabilising influence of the flow-mediated interactions that promotes orderly positioning can be overturned by a global destabilising effect that results from many-body effects. Recent studies have reported that a new type of self-amplifying wave, termed ``flonon,'' can cause collisions between flyers in groups consisting of multiple individuals~\cite{newbolt2024flow,nitsche2025stability,mavroyiakoumou2025modeling}.

Therefore, it is important to understand the extent to which the flow-mediated interactions provide aerodynamic benefits to members of flying groups, but also the level of sensing and feedback control mechanisms that individual flyers need to use to maintain group cohesion. 
Previous work considered a feedback control mechanism based on local flow sensing to stabilise a pair of frequency-uncoordinated swimmers into a cohesive formation~\cite{hang2024flow} since the stability of the in-line formations is sensitive to finite mismatch in frequencies~\cite{newbolt2019flow}.
Other works have introduced an active mechanism in longer arrays of flyers, in which the individuals adjust their flapping amplitude based on their propulsive speed relative to their upstream neighbour to avoid collisions~\cite{nitsche2025stability}. In these examples, the flapping motion is a sinusoidal function with kinematic parameters (amplitude or frequency) that are no longer constants, but rather functions of time, dynamically determined through the flow-interaction forces.

The main objective of the current work is to study flapping strategies for flyers moving in an in-line arrangement, whose goal is to maintain a constant distance between them. Therefore, in this case, we investigate a different scenario in which the horizontal distance between the self-propelling flyers is a prescribed constant, and the flapping (vertical) motion of the flyers is now an outcome of the flow-mediated interactions, given a flapping motion for the leader in the group. To address this, we use the simple mathematical model, developed in our previous work, that describes the general features of collective and interactive locomotion dynamics and can easily scale up to arbitrarily large groups~\cite{mavroyiakoumou2025modeling}. It is a unified framework that has been shown to capture all prior experimental results of flapping foils within a single model of formation flight. The model is an improved version of previous aerodynamic models that describe the fluid-structure interactions among the flyers through a simplified representation of the wake flow generated by their flapping motions~\cite{newbolt2019flow,newbolt2024flow}.

The reach of our study extends beyond understanding the role of flow interactions in flying and swimming formations. It has applications in UAV (unmanned air vehicle) formation control, in which agents move in such a way as to maintain a fixed distance between them~\cite{anderson2008uav}. The formation of UAVs in a sense moves as a rigid entity in the direction of propulsion and previous studies focused on characterizing the choice of agent pairs to secure this shape-preserving property, using leader-follower control~\cite{rafifandi2019leader,ali2021leader,mercado2013quadrotors,ghamry2015formation,chen2023leader}. The results of our current work can also be applied to autonomous underwater vehicles or robotic fish swarms~\cite{yoon2010cooperative}.

The structure of the paper is as follows. In \S\ref{sec:dynamicalModelDDEs} we describe our follower-wake interaction model formulated using a system of delay differential equations. Using a new, general framework, we reformulate the aerodynamical model using a system of iterative ordinary differential equations (\S\ref{sec:dynamicalModelODEs}), which makes the model more amenable to analysis. In \S\ref{sec:asymptotics}, we derive detailed asymptotic approximations for the case of a single, isolated flyer flapping with a prescribed sinusoidal function. In \S\ref{sec:strategies} we derive the flapping strategies flyers in a group should adopt to maintain a constant distance from their upstream neighbour, given a flapping motion for the leader. By employing efficiency and stability arguments, we also determine whether flyers, whose goal is to maintain a constant separation between them, should choose to flap in phase or out of phase with their upstream neighbours (\S\S\ref{sec:inphase}--\ref{sec:StabilityMainText}).
In \S\ref{sec:higherModes}, we consider the stabilisation dynamics of flyers in a group when the leader is flapping with an arbitrary periodic velocity. Section~\ref{sec:conclusions} presents the conclusions.

\section{Flyer-wake interaction model}\label{sec:model}

We begin by presenting a dynamical model of collective locomotion based on fluid-mechanical interactions between flyers that are moving in a semi-infinite quiescent fluid~\cite{mavroyiakoumou2025modeling}. We also non-dimensionalise the governing equations and find the main dimensionless parameters that dictate the group dynamics.

\subsection{Model formulation as a system of delay differential equations}\label{sec:dynamicalModelDDEs}
\begin{figure}
    \centering
    \includegraphics[width=.98\textwidth]{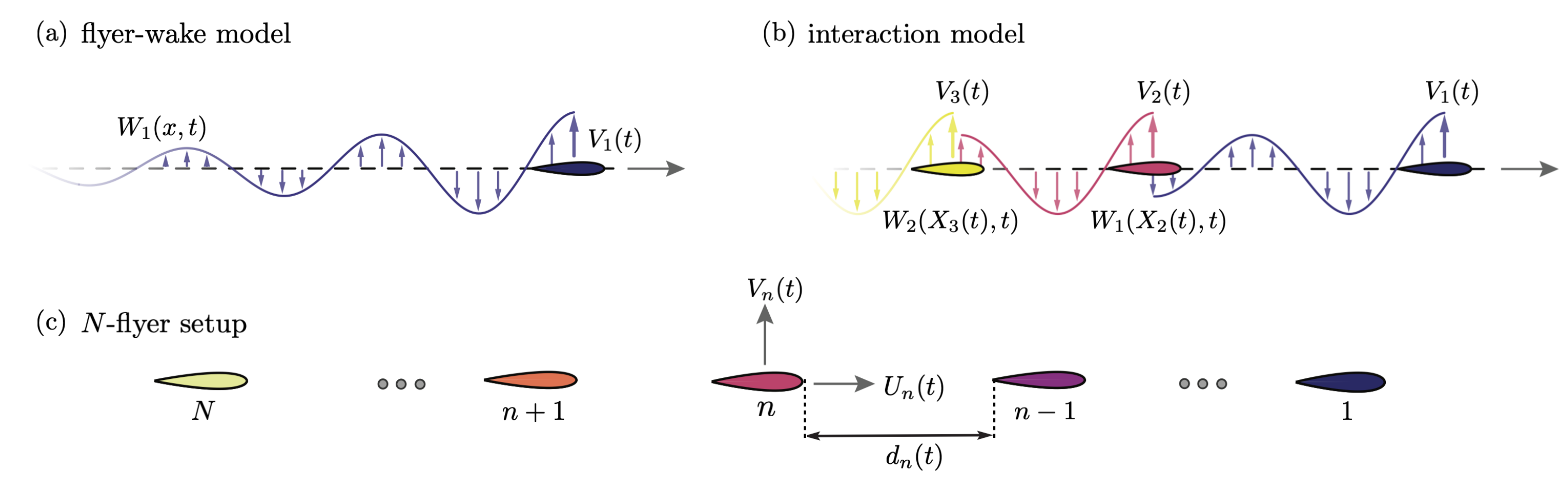}
    \caption{Schematic diagrams demonstrating the flyer-wake model. (a) Each flyer generates a wake signal with speed $W_1(x,t)$, directly related to its flapping velocity $V_1(t)$, with the wake flow thereafter decaying in time. (b) An interaction model used to simulate the group dynamics with one-way interactions. The next flyer (i.e.\ the follower) experiences a thrust force that depends on its flapping velocity~$V_2(t)$ relative to the ambient wake $W_1(X_2(t),t)$ at position $x=X_2(t)$. (c)~Problem setup of a linear formation of $N$ flapping flyers indexed by $n$, each undergoing heaving-and-plunging motions with velocity $V_n(t)$ and free forward flight motions with speed~$U_n(t)$. Flyer $n$ and its upstream neighbour $n-1$ are separated by a distance $d_n(t)=X_{n-1}(t)-X_n(t)\geq 0$.}
    \label{fig:modelSchem}
\end{figure}

The model presented in this section is the same as the one developed in our previous work~\cite{mavroyiakoumou2025modeling}, but we repeat it briefly for completeness. Specifically, the aerodynamic model involves flyers moving along a line, creating a wake flow signal and interacting with the wake flows generated by others. Each flyer's flapping motion determines both its self-propulsion dynamics and the wake flow signal left behind in its trail. This is shown schematically in figure~\ref{fig:modelSchem}(a). 
The propulsive force that the flyer experiences is dependent upon the interaction between its instantaneous flapping signal and the ambient wake produced by other individuals in the linear flying formation (see figure~\ref{fig:modelSchem}(b)). The flow interactions are  restricted to nearest neighbours and are also one-way in the downstream direction, implying that within a pair of flyers the follower is influenced by the leader, but not vice versa~\cite{mavroyiakoumou2025modeling,newbolt2024flow}. In our framework, interactions occur via long-lived flows that retain memory of the conditions under which they were created~\cite{becker2015hydrodynamic,mavroyiakoumou2025modeling}.

Each flyer $n$ is treated as an inertial body that is free to move in the horizontal direction due to aerodynamic interactions, and has instantaneous horizontal position~$X_n(t)$, propulsive speed $U_n(t)=\dot{X}_n(t)$, and flapping velocity $V_n(t)$, the latter representing vertical oscillations (heaving-and-plunging motions). The indexing ${n=1,2,\dots, N}$ is used to label each of the flyers in the group, with $n=1$ representing the leader and $n=N$ the last member, as shown in figure~\ref{fig:modelSchem}(c).
Each individual's horizontal dynamics are governed by Newton's second law of motion, as follows:
\begin{equation}\label{eq:N2L}
    M_n\dot{U}_n(t)=T_n(t)-D_n(t),\quad n=1,2,\dots, N,
\end{equation}
where $M_n$ is the mass of flyer $n$, $\dot{U}_n(t)$ is the acceleration in the horizontal direction, $T_n(t)$ is the horizontal thrust force, and $D_n(t)$ is a skin-friction drag. In this work, we assume that all flyers have equal mass,~$M_n=M$, for simplicity. In the absence of any interactions, as in the case of a single isolated flyer, thrust varies quadratically with the vertical flapping velocity, $\rho C_TcsV_n^2/2$, and drag varies as the $3/2$-power of the horizontal propulsion speed, $C_Ds\sqrt{\rho\mu c}\, U_n^{3/2}/2$~\cite{floryan2017scaling,smits2019undulatory,heydari2021school}. The skin friction drag along the upper and lower flyer surfaces is modelled using Blasius laminar boundary layer theory~\cite{schlichting2016boundary} and is derived through $\rho  (C_D/\sqrt{\mathrm{Re}})cs U_n^{2}/2$, where $\mathrm{Re}=\rho U_nc/\mu$ is  the Reynolds number. Details about the choice of drag law power can be found in our previous work~\cite{mavroyiakoumou2025modeling,fang2025flowinteractionsforwardflight}. These aerodynamic forces involve the fluid density $\rho$, the dynamic viscosity $\mu$, the wing's chord and span length $c$ and $s$, and the dimensionless coefficients of thrust and drag, $C_T$ and $C_D$, respectively~\cite{tritton2012physical,vandenberghe2006unidirectional,vandenberghe2004symmetry}.

A relative flow
model, in which the thrust force is adjusted to depend on the vertical flapping velocity of the flyer relative to an ambient wake signal $\Delta V_{n}(t)$, is used to capture the aerodynamic interactions~\cite{mavroyiakoumou2025modeling}. The wake created by flyer $n$ at position $x$ and time $t$ has flow speed $W_n(x,t)$. We assume that, at the flyer's location $x=X_n(t)$, this flow speed is equal to the flapping velocity ${W_{n}(X_n(t),t)=V_{n}(t)}$~\cite{becker2015hydrodynamic,ramananarivo2016flow,newbolt2019flow,oza2019lattices,newbolt2024flow}. In the model, the thrust force experienced by flyer $n$ is proportional to the square of $\Delta V_{n}(t)=V_{n}(t)-W_{n-1}(X_{n}(t),t)$. The model also incorporates an ``erase-and-replace'' scheme, meaning that each flyer overwrites the wake signal of its upstream neighbour with its own, as shown in the schematic diagram in figure~\ref{fig:modelSchem}(b). 
The flyer located at $X_{n-1}(t_n(t))$ creates a wake signal with flow velocity given by
${W_{n-1}(X_n(t),t)=V_{n-1}(t_n(t))e^{-(t-t_n(t))/\tau}}$, and, specifically, this signal decays exponentially in time with dissipation timescale~$\tau$~\cite{ramananarivo2016flow,newbolt2019flow}. We note that these terms involve $t_n(t)$, which corresponds to the earlier time when the upstream neighbour $n-1$ was at the current position of flyer $n$ at the current time $t$, as defined implicitly via $X_{n-1}(t_n(t))=X_n(t)$. Therefore, the effect of memory in the system is encoded in the variable $t_n(t)$, while the delay between when a signal left by flyer $n-1$ is encountered by flyer $n$ is given by the difference $t-t_n(t)$. 
Since the memory timescale $t_n(t)$ depends on time itself, we include it as a state variable in the dynamical system. We derive its evolution equation using the chain rule: $\dot{X}_{n-1}(t_n(t))\dot{t}_n(t)=\dot{X}_n(t)$, which yields ${\dot{t}_n(t)=\dot{X}_n(t)/\dot{X}_{n-1}(t_n(t))=U_n(t)/U_{n-1}(t_n(t))}$, with the denominator $U_{n-1}(t_n(t))$ constituting a state-dependent delay.

Therefore, to determine the horizontal propulsion dynamics of each flyer $n$, we solve the following system of nonlinear delay differential equations for $(X_n, U_n,t_n)$:
\begin{subequations}
    \begin{align}
    \dot{X}_n(t)&=U_n(t),\label{eq:xdot}\\
    \dot{U}_n(t)&=\frac{\rho C_Tcs}{2M}\left[V_n(t)-V_{n-1}(t_n(t))e^{-(t-t_n(t))/\tau}\right]^2-\frac{C_Ds\sqrt{\rho \mu c}}{2M} U_n^{3/2}(t),\label{eq:UndotDimensional}\\
    \dot{t}_n(t)&=\frac{U_n(t)}{U_{n-1}(t_n(t))}.\label{eq:tnsystem}
\end{align}
\end{subequations}
For a group consisting of $N$ flyers, we solve $3N$ such equations.

\subsection{Model reformulation as a system of ordinary differential equations}\label{sec:dynamicalModelODEs}

The system of delay differential equations presented in \S\ref{sec:dynamicalModelDDEs} can be reformulated as a system of ordinary differential equations that are solved iteratively, and this formulation of the propulsion dynamics is more amenable to analysis. 
In this work, the ensemble is flying in a semi-infinite domain of quiescent fluid and so the leader flies as a solo flyer, in the absence of any flow interactions. Therefore, the interaction term involving $V_{n-1}$ is removed from~\eqref{eq:UndotDimensional}, as is~\eqref{eq:tnsystem} entirely. 

In \S\ref{sec:dynamicalModelDDEs}, we solve for a pair $(X_n(t), U_n(t))$ for each flyer $n$. The idea now is to express the position $x=X_n(t)$ as an inverse map given by $t=X_n^{-1}(x)$. This is possible because the instantaneous horizontal positions $X_n(t)$ are bijective functions, so their inverse $X_n^{-1}(x)$ exists. Therefore, the temporal representation can be written in terms of spatial coordinates instead, as follows:
\begin{equation}
    (X_n(t),U_n(t))\to (x,U_n(X_n^{-1}(x))).
\end{equation}

The definition of $t_n(t)$ is $X_{n-1}(t_n(t))=X_n(t)$, and if we determine what this means in terms of the wake velocity $W_{n-1}(x=X_n(t),t)$, then we can simplify the wake velocity at the position of flyer~$n$, given by $V_{n-1}(t_n(t))e^{-(t-t_n(t))/\tau}$. To do that, we note that:
\begin{equation}
    t_n(t)=X_{n-1}^{-1}(X_n(t)).
\end{equation}
Therefore, we have
\begin{equation}
    W_{n-1}(x=X_n(t),t)=V_{n-1}(t_n(t))e^{-(t-t_n(t))/\tau}=V_{n-1}(X_{n-1}^{-1}(X_n(t)))e^{-\left(t-X_{n-1}^{-1}(X_n(t))\right)/\tau}.
\end{equation}
This implies that the system of delay differential equations (Eqs.~\eqref{eq:xdot}--\eqref{eq:tnsystem}) can be written as ordinary differential equations, where the leader is treated separately (as before), and the rest of the flyers are determined iteratively. For the leader, we solve for: 
\begin{subequations}\label{eq:leaderDiffEqn}
\begin{align}
\dot{X}_1(t)&=U_1(t),\\
\dot{U}_1(t)&=\frac{\rho C_T cs}{2M}V^2_1(t)-\frac{C_D s\sqrt{\rho\mu c}}{2M} U_1^{3/2}(t).
\end{align}
\end{subequations}
Therefore, for the remaining flyers $n>1$, the position $x=X_{n-1}(t)$ will be obtained as part of the solution of the dynamical equations for the upstream neighbour $n-1$, from which the inverse map $t=X_{n-1}^{-1}(x)$ can then be computed numerically through interpolation. This allows us to iteratively solve for each flyer $n$ through:
\begin{subequations}\label{eq:EquationsMotionDimensional}
\begin{align}
\dot{X}_n(t)&=U_n(t),\label{eq:odeX}\\
\dot{U}_n(t)&=\frac{\rho C_T cs}{2M}\left[V_n(t)-V_{n-1}(X_{n-1}^{-1}(X_n(t)))e^{-\left(t-X_{n-1}^{-1}(X_n(t))\right)/\tau}\right]^2-\frac{C_D s\sqrt{\rho\mu c}}{2M} U_n^{3/2}(t).\label{eq:odeU}
\end{align}
\end{subequations}
To solve the full equations of motion \eqref{eq:leaderDiffEqn}--\eqref{eq:EquationsMotionDimensional} numerically, we use the high-order Runge-Kutta method \texttt{DOP853} available in \texttt{SciPy}'s \texttt{solve\_ivp} solver
in Python~\cite{hagen2019class}.

\subsection{Non-dimensionalisation}\label{sec:modelDimensionless}

Before analyzing the system of flyers, we recast the model shown in Sec.~\ref{sec:dynamicalModelODEs} in dimensionless form.
The governing equations~\eqref{eq:odeX}--\eqref{eq:odeU} are non-dimensionalised by the characteristic timescale~$1/f$, the period-averaged equilibrium speed of a solo flyer $U^*$, and the typical flapping speed $\pi Af$. In particular, we take
\begin{align}
    \tilde t =f t,\qquad \tilde U_n =\frac{U_n}{U^*},\qquad \tilde X_n =\frac{X_nf}{U^*},\qquad \tilde V_n(\tilde t) =\frac{V_n(t)}{\pi Af},\qquad \text{with } U^*=\left(\frac{\pi^2 C_TA^2 f^2}{2C_D}\sqrt{\frac{\rho c}{\mu}}\right)^{2/3}.
\end{align}

With these dimensionless variables,~\eqref{eq:odeU} becomes
\begin{equation}\label{eq:nondimensionalizingUndot}
    fU^*\dfrac{\d \tilde{U}_n}{\d \tilde{t}}=\frac{\rho C_T cs(\pi  A f)^2}{2M}\left[\tilde{V}_n(\tilde{t})-\tilde{V}_{n-1}(\tilde{t}_n(\tilde{t}))e^{-\left(\tilde t-\tilde t_n(\tilde t)\right)/(\tau f)}\right]^2-\frac{C_Ds\sqrt{\rho \mu c}}{2M}(U^*)^{3/2}\tilde{U}_n^{3/2},
\end{equation}
with dimensionless quantities (and their dimensionless derivatives) denoted by tildes.
Dividing~\eqref{eq:nondimensionalizingUndot} by $fU^*$, the dimensionless equations of motion for each flyer $n$ become
\begin{subequations}
\begin{align}
    \dot{\tilde X}_n &= \tilde U_n,\label{eq:XndotDimensionless}\\
    \dot{\tilde{U}}_n &= \gamma \left(2 \tilde B_n^2-\tilde U_n^{3/2}\right),\label{eq:UndotDimensionless}\\
    \tilde B_n &=\tilde V_n(\tilde t)-\tilde V_{n-1}(\tilde X_{n-1}^{-1}(\tilde X_n(\tilde t)))e^{-\left(\tilde t-\tilde X_{n-1}^{-1}(\tilde X_n(\tilde t)\right)/\chi},
\end{align}
\end{subequations}
with well-posed initial conditions
\begin{align}\label{eq:ICs}
    \tilde X_n(0)=-(n-1)\tilde d_0, \qquad \tilde U_n(0)=\tilde U_0. 
\end{align}
The dimensionless groups are 
\begin{align}\label{eq:dimensionlessGroups}
    \gamma =\frac{s (C_T \mu/f )^{1/3} ( \pi C_D A \rho c)^{2/3}}{2^{4/3} M},\qquad \tilde d_0=\frac{d(0) f}{U^*},\qquad  \chi=\tau f, \qquad C_T,\qquad C_D,
\end{align}
where $C_T$ is the coefficient of thrust, and $C_D$ is the coefficient of skin friction drag, set as $C_T=1$ and $C_D=10$, respectively. In our previous work~\cite{mavroyiakoumou2025modeling}, we showed that, when $C_T$ is fixed at 1, both for simplicity and to be consistent with experiments~\cite{newbolt2019flow,newbolt2024flow}, a value of $C_D=10$ is necessary to reproduce  experimental results. Here $\chi$ characterises how the timescale over which the wake dissipates compares to the flapping period.
Representative values of the dimensionless groups using parameter values from Table~\ref{tab:ParameterValues}, relevant to experiments with flapping foils in water~\cite{becker2015hydrodynamic,ramananarivo2016flow,newbolt2019flow,newbolt2024flow}, are $\gamma \approx 1.3$ and $\chi=1.25$, with $U^* \approx 68 \,\text{cm s}^{-1}$. Throughout this work, unless otherwise stated, we will use $\chi=1$. 

The main dimensionless group dictating the propulsion dynamics is $\gamma$, capturing the physical dimensions and kinematic parameters of the flyers, as well as the fluid parameters. It can be expressed in terms of other known dimensionless numbers as follows:
\begin{align}
    \gamma =\frac{(C_D^2C_T)^{1/3}\mathrm{Re}_f^{2/3}}{\mathrm{Wo}^{2}M^*}\left(\frac{\pi ^5 }{2}\right)^{1/3},\qquad \text{ with }\mathrm{Re}_f=\frac{\rho A f c}{\mu},\;\; M^*=\frac{M}{\rho c^2 s},\;\; \mathrm{Wo}=c\left(\frac{2\pi f}{\nu}\right)^{1/2},
\end{align}
where $\mathrm{Re}_f$ is the flapping Reynolds number, $M^*$ is the dimensionless mass, and Wo is the Womersley number with $\nu=\mu/\rho$ the kinematic viscosity of the fluid. The parameter $\gamma$ corresponds to the self-generated thrust from the flapping (see \S\ref{sec:connectionEqmDist} for details), and is a measure of the effectiveness of flapping motion in generating aerodynamic forces relative to inertial effects. Specifically, $\gamma$ captures the balance between unsteady aerodynamic loading and the combined structural and fluid inertia that resists the motion. A high $\gamma$ indicates that flapping can produce aerodynamic forces efficiently because inertial resistance is comparatively weak. This is consistent with light flyers with slow wingbeats. Conversely, a low $\gamma$ corresponds to inertia-dominated regimes in which flapping is less effective, typical of heavier flyers or small-amplitude and fast wingbeats.

For ease of notation, in the remaining sections the tildes are dropped from the dimensionless quantities.

\section{Asymptotic approximation of propulsive speed of a single flyer}\label{sec:asymptotics}
In this section we construct an asymptotic approximation to the emergent propulsive speed $U_1$ that will prove useful later. We consider an isolated flyer with flapping velocity $V_1(t)=\cos(2\pi t)$, and whose emergent propulsive velocity obeys $\dot U_1(t)=\gamma\left(2 V_1^2(t)- U_1^{3/2}(t)\right)$.

We first treat the regime of small $\gamma$, which is the range most relevant to previous experiments. When the period-averaged speed is $u^\ast=1$, we write $U_1(t)=u^\ast + w(t)$ and linearise for small $w(t)$, obtaining
\begin{align}
    \dot{w}+\frac{3\gamma}{2}w=\gamma\cos(4\pi t),
\end{align}
whose transient-free solution is
\begin{align}
    \label{eq:SingleFlyerFirstOrderVelocity}
    w(t)=R(\gamma)\cos(4\pi t+\varphi(\gamma)),\qquad
    R(\gamma)=\frac{2}{3}\frac{\gamma}{\sqrt{\gamma^2+16\pi^2/9}},\qquad
    \varphi(\gamma)=-\tan^{-1}\!\left(\frac{8\pi}{3\gamma}\right).
\end{align}
In appendix~\ref{sec:AppSingleFlyer} we extend this result and derive higher order corrections. The expansion there yields
\begin{subequations}
    \begin{align}
    &U_1(t)=1+C + A_0e^{i\omega t}+A_0^*e^{-i\omega t}+Be^{2i\omega t}+B^*e^{-2i\omega t},
    \\
    &A_0 = \frac{\gamma/2}{3\gamma/2 + i\omega},\quad     B =-\frac{3\gamma}{8}\,
\frac{A_0^{\,2}}{3\gamma/2 + i\,2\omega},\quad 
    C = -\frac12\,|A_0|^2=-\frac{R^2}{8},\label{eq:A0BandC}
\end{align}
\end{subequations}
where $\omega=4\pi$. We make particular note of the constant correction $C(\gamma)<0$, which lowers the period averaged speed at second order. In appendix~\ref{sec:AppSingleFlyer} we compare these asymptotic expressions with numerical solutions and show good agreement for the parameter ranges of interest. The frequency analysis there also shows higher order Fourier coefficients decay rapidly, which supports the validity of the first-order approximation for moderate $\gamma$. The correction $C=-R^2/8$ will be relevant later when we study interactions between consecutive flyers. 

We use \eqref{eq:SingleFlyerFirstOrderVelocity} to initialise the numerical solution of \eqref{eq:EquationsMotionDimensional} in such a way that transient effects are minimised. In particular, choosing $U_1(0)=1+w(0)$, we force the system to have no first order transients, converging to the steady-state solution extremely quickly and limiting the number of periods we have to simulate before we consider the numerical solution to be transient-free.

For larger values of $\gamma$ the response becomes quasi-steady. The leading-order solution then follows from a balance between thrust and drag:
\begin{align}\label{eq:LeadingOrderVelocityLargeGamma}
    U_1(t)\approx (2\cos^2(2\pi t))^{2/3}+O(\gamma^{-1})=\left(1+\cos(4\pi t)\right)^{2/3}+O(\gamma^{-1}),\qquad \gamma\to\infty. 
\end{align}
Hence, in the large-$\gamma$ limit the maximum speed is $\max_t U_1(t)=2^{2/3}\approx 1.59$.

To produce an approximation that is accurate across both limits, we combine the small and large $\gamma$ results to obtain
\begin{align}\label{eq:CombinedApproximation}
    U_1(t)=\left(1+\frac{3}{2}R(\gamma)\cos(4\pi t+\varphi(\gamma))\right)^{2/3}.
\end{align}
For $\gamma\ll1$ we have $R\ll1$ and the above expands to $1+R\cos(4\pi t+\varphi)$, which recovers the small-$\gamma$ result. For $\gamma\gg1$ we have $R\to 2/3$ and $\varphi\to0$, and the expression reduces to the quasi-steady form in \eqref{eq:LeadingOrderVelocityLargeGamma}. In appendix~\ref{sec:AppSingleFlyer} we compare \eqref{eq:CombinedApproximation} with numerical solutions and find good agreement over the intermediate parameter range as well.

The asymptotic solution above provides a convenient initial condition for time integration of the full system, since it captures the main oscillatory content and thus reduces large transients in numerical simulations.

\section{Enforcing constant distance between flyers}\label{sec:strategies}

We now consider the problem of determining the flapping velocities $V_n(t)$ required to maintain a prescribed constant separation distance between consecutive flyers. Specifically, we impose
\begin{align}
    d_n(t)=X_{n-1}(t)-X_n(t)=d,
\end{align}
for all times $t\geq0$ and for all flyers $n=2,3,\dots,N$.

Maintaining a fixed separation may be advantageous for several reasons. From a biological perspective, it prevents collisions while still allowing individuals to benefit from flow-mediated aerodynamic interactions. Field observations often indicate that birds within flocks maintain an approximately constant nearest-neighbour distance~\cite{gould1974vee,major1978three}. In schooling fish, swimming too close may increase vulnerability to predation, for example during engulfment feeding by baleen and humpback whales~\cite{goldbogen2017baleen,cade2020predator}, while swimming too far apart reduces hydrodynamic benefits. A fixed separation may therefore reflect a balance between energetic efficiency and risk mitigation. From an engineering standpoint, maintaining constant spacing is a central objective in the control of UAV and autonomous vehicle formations~\cite{anderson2008uav}.

In this work, constant separation is enforced by allowing each flyer to adjust its flapping kinematics. The leader of the group is unconstrained and flaps according to a prescribed vertical velocity $V_1(t)$. All downstream flyers adapt their flapping velocities so as to maintain a distance $d$ from their immediate upstream neighbour.

\subsection{Flapping strategies that enforce constant separation}
Suppose that each flapping flyer is free to choose its vertical velocity $V_n(t)$.  
Its motion evolves according to
\begin{subequations}
\begin{align}
    \dot X_n(t) &= U_n(t),\\
    \dot U_n(t) &= \gamma\!\left(2B_n^2(t) - U_n^{3/2}(t)\right),\label{eq:undotopen}
\end{align}
\end{subequations}
where the \emph{effective} thrust experienced by flyer $n$ is given by
\begin{align}
    B_n(t)=
    \begin{cases}
        V_1(t), & n=1,\\
        V_n(t)
        - V_{n-1}\!\left(X_{n-1}^{-1}(X_n(t))\right)
          e^{-\left(t - X_{n-1}^{-1}(X_n(t))\right)/\chi}, & n>1.
    \end{cases}
\end{align}
We assume that the objective of each follower is to maintain a constant spacing $d$ relative to its upstream neighbour. Since the leader is unaffected by others, it may choose any flapping velocity $V_1(t)$. In contrast, all followers must match the leader's propulsive speed in order to preserve spacing. This implies
\begin{align}
    U_n(t)=U_1(t), \qquad \text{for }n=2,3,\dots,N,
\end{align}
and hence
\begin{align}\label{eq:xnspacing}
    X_n(t)=X_1(t)-(n-1)d.
\end{align}

Substituting this condition into \eqref{eq:undotopen} yields
\begin{align}
    2\gamma B_n^2(t)
    &= \dot U_n + \gamma U_n^{3/2}
      = \dot U_1 + \gamma U_1^{3/2}
      = 2\gamma B_1^2(t),
\end{align}
which implies that $B_n^2(t)=B_1^2(t)$. Therefore, the flapping velocities must satisfy
\begin{align}
    V_n(t)
    - V_{n-1}\!\left(X_{n-1}^{-1}(X_n(t))\right)
      e^{-\left(t - X_{n-1}^{-1}(X_n(t))\right)/\chi}
    = \pm V_1(t).
\end{align}
This relation determines the flapping velocity of each follower recursively:
\begin{align}
    V_n(t)
    = \pm V_1(t)
      + V_{n-1}\!\left(t_n(t)\right)
        e^{-\left(t - t_n(t)\right)/\chi},
\end{align}
where $t_n(t)=X_{n-1}^{-1}\!\big(X_n(t)\big)$.

Using the enforced spacing \eqref{eq:xnspacing}, we observe that
\begin{align}\label{eq:xnminus1tn}
    X_{n-1}\!\left(t_n(t)\right)
    &= X_1\!\left(t_n(t)\right)-(n-2)d,
\end{align}
and since $X_{n-1}\!\left(t_n(t)\right)=X_n(t)=X_1(t)-(n-1)d$, it follows that
\begin{align}
    X_1\!\left(t_n(t)\right)=X_1(t)-d.
\end{align}
Consequently,
\begin{align}
    t_n(t)=t_2(t)=X_1^{-1}\!\left(X_1(t)-d\right),
\end{align}
so that all followers experience the same time-shifted wake signal determined solely by the leader's trajectory.

The separation-preserving flapping rule therefore reduces to
\begin{align}\label{eq:RecursiveFormulainTermsoft_2}
    V_n(t)
    = \sigma_n V_1(t)
      + V_{n-1}\!\left(t_2(t)\right)
        e^{-\left(t-t_2(t)\right)/\chi},
    \qquad
    t_2(t)=X_1^{-1}\!\left(X_1(t)-d\right),
\end{align}
where $\sigma_n=\pm1$ specifies the sign choice for flyer $n$.

Equation~\eqref{eq:RecursiveFormulainTermsoft_2} is the central result of this section. It provides the flapping velocity required for each flyer to maintain a prescribed separation from its upstream neighbour, given the leader's flapping motion.

\subsection{Arbitrary leader flapping velocity}

The recursive formula \eqref{eq:RecursiveFormulainTermsoft_2} may be iterated to express the flapping velocity of any follower explicitly in terms of the leader. For example,
\begin{align}\begin{split}
    V_2(t)&=\sigma_2 V_1(t)+V_1(t_2(t))e^{-t/\chi}e^{t_2(t)/\chi},\\
    V_3(t)&=\sigma_3 V_1(t)+\left[\sigma_2 V_1(t_2(t))+V_1(t_2(t_2(t)))e^{-t_2(t)/\chi}e^{t_2(t_2(t))/\chi}\right]e^{-t/\chi}e^{t_2(t)/\chi}.
    \end{split}
\end{align}
If we consider then the earlier time $t_n(t)$, we obtain recursively that
\begin{align}\begin{split}
    t_3(t)&\equiv t_2(t_2(t))=X_1^{-1}(X_1(t_2(t))-d)=X_1^{-1}(X_1(X_1^{-1}(X_1(t)-d))-d)=X_1^{-1}(X_1(t)-2d),\\
    t_4(t)&\equiv t_2(t_2(t_2(t)))=X_1^{-1}(X_1(t)-3d),\\
    t_n(t)&=t_2(t_2(\dots))=X_1^{-1}(X_1(t)-(n-1)d),
    \end{split}
\end{align}
(and $t_1(t)\equiv t$). This allows the recursive relation \eqref{eq:RecursiveFormulainTermsoft_2} to be written compactly as
\begin{align}
    V_n(t)=\sum_{j=1}^n\sigma_j V_1(t_{n-j+1}(t))e^{-t/\chi}e^{t_{n-j+1}(t)/\chi}=\sum_{j=1}^n \sigma_{n-j+1}V_1(t_j(t))e^{-t/\chi}e^{t_j(t)/\chi},
\end{align}
for $n\geq 2$.
The following closed form of $V_n(t)$ can be obtained for flyer $n$, expressed in terms of the position of the first flyer $X_1(t)$:
\begin{align}\label{eq:GeneralFormulaVn}
    \begin{split}V_n(t)&=\sum_{j=1}^n\sigma_{n-j+1}V_1(X_1^{-1}(X_1(t)-(j-1)d))e^{-t/\chi}e^{X_1^{-1}(X_1(t)-(j-1)d)/\chi}\\&=\sum_{j=0}^{n-1}\sigma_{n-j}V_1(X_1^{-1}(X_1(t)-jd))e^{-t/\chi}e^{X_1^{-1}(X_1(t)-jd)/\chi}.
    \end{split}
\end{align}
We now specialise to the case where the leader flaps sinusoidally with $V_1(t)=\cos(2\pi t)$. Two natural sign choices arise. The first corresponds to \emph{in-phase} flapping, where $\sigma_n=+1$ for all $n$. The second corresponds to \emph{out-of-phase} flapping, where $\sigma_n=(-1)^{n+1}$.

In the following sections we analyse these two strategies in detail and assess their energetic cost and dynamical stability.

\section{Flapping in or out of phase with upstream neighbour}

In this section we study separately the case of in-phase and out-of-phase flapping for the trailing flyers (which corresponds to a sign choice of $\sigma_n$), given that the leader in the group flaps sinusoidally with $V_1(t)=\cos(2\pi t)$. We also carry out a stability analysis to determine which of the two choices is preferable and make a connection with stable emergent equilibrium distances studied in previous works~\cite{mavroyiakoumou2025modeling}.

\subsection{Flapping in phase}\label{sec:inphase}
\begin{figure}[htpb!]
    \centering
    \includegraphics[width=\linewidth]{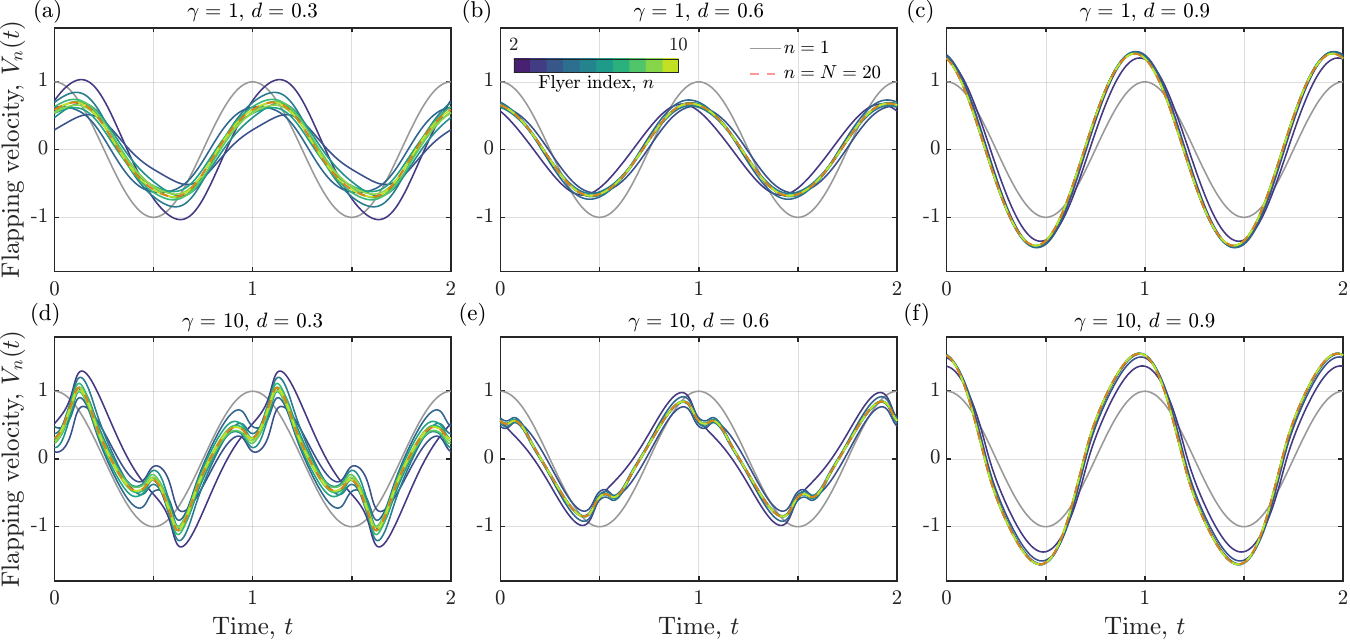}
    \caption{Flapping velocities $V_n(t)$ for each flyer $n=2,3,\dots, N$ to maintain a constant distance $d$ (equal to 0.3, 0.6, and 0.9; increasing from left to right) with its upstream neighbour for $\gamma=1$ (panels (a)--(c)) and $\gamma=10$ (panels (d)--(f)), given that the leader in the group flaps sinusoidally with velocity $V_1(t)=\cos(2\pi t)$ (grey curves). Here the group size is $N=20$ and $\chi=1$. The flapping velocities of flyers $n=2$ to 10 are shown in shades of blue to green and the last member in the group $n=N=20$ is displayed as a dashed red curve.}
    \label{fig:Vn_gamma_d_inphase}
\end{figure}

For a leader that flaps sinusoidally with $V_1(t)=\cos(2\pi t)$, we analyse the flapping velocities when each follower remains \emph{in phase} with its upstream neighbour, i.e. $\sigma_n=+1$. In figure~\ref{fig:Vn_gamma_d_inphase} we plot numerically computed $V_n(t;\gamma,d)$ for two values of $\gamma$ (1 and 10; increasing from top to bottom) and three values of $d$ ($0.3,0.6,0.9$; increasing from left to right), with $\chi=1$. Both $\gamma$ and $d$ strongly influence the shape and amplitude of $V_n(t)$. For this choice of $\sigma_n=+1$, the $L^2$ norm $\|V_n(t)\|_2$ is small for $d=0.6$ (figures~\ref{fig:Vn_gamma_d_inphase}(b) and \ref{fig:Vn_gamma_d_inphase}(e)) and large for $d=0.9$ (figures~\ref{fig:Vn_gamma_d_inphase}(c) and~\ref{fig:Vn_gamma_d_inphase}(f)), suggesting that it is significantly costlier for the flyers to maintain a separation of $d=0.9$. By contrast, the more compact configuration with $d=0.6$ is more energy efficient than solo flying, since $\|V_n(t)\|_2\approx0.6<1$, indicating an aerodynamic benefit from the formation.

For small and moderate values of $\gamma$ the emergent propulsive velocity is approximately constant, so that the flapping velocities $V_n(t)$ are close to cosines. The amplitude of these cosines is strongly influenced by the separation distance. For larger $\gamma$ and small $d$ (which corresponds to strong aerodynamic interactions) the flapping velocities contain non-negligible higher-order Fourier modes. In particular, the third and fifth modes grow in importance as $\gamma$ increases, while even modes remain negligible even for very large $\gamma$. A detailed frequency analysis is presented in appendix~\ref{sec:AppFFTinPhase}, where we also rationalise the appearance of higher-order frequencies for $\gamma\gg1$.  

For all parameter values considered, the flapping functions converge to a limiting profile $V_\infty(t)$ as $n\to\infty$ (plotted in red in figure~\ref{fig:Vn_gamma_d_inphase}). Evidence for this exponential convergence in $n\tilde d/\chi$ is provided in appendix~\ref{sec:AppConvergenceInN}. We now compute $V_\infty(t)$ analytically for small and moderate values of $\gamma$.

In appendix~\ref{sec:app_details_t_2} we show that to leading order in $R$, $t_2(t)=t-\tilde d$, where $\tilde d=d/(1+C)$ and $C$ is given in~\eqref{eq:A0BandC}. From the definition in \eqref{eq:SingleFlyerFirstOrderVelocity}, we see that this approximation is valid for $\gamma\lesssim1$, where $R\ll1$. Expanding $V_n(t)=V_n^{(0)}(t)+R\,V_n^{(1)}(t)+\dots$, the leading-order flapping velocity is
\begin{align}
    V_n^{(0)}(t)=\sum_{j=0}^{n-1}\cos\left(2\pi\left[t-j\tilde d\right]\right)e^{-\tilde d j/\chi}+O(R).
\end{align}
Evaluating the finite geometric sum yields
\begin{gather}
    V_n^{(0)}(t)=\frac{1}{2}\left(J_ne^{2\pi i t}+J_n^\ast e^{-2\pi i t}\right),\qquad
    J_n=\frac{1-e^{-\tilde d\, n\,\left({1}/{\chi }+2\pi i\right)}}{1-e^{-\tilde d\,\left({1}/{\chi}+2\pi i\right)}},\nonumber\\ 
    \vert J_n\vert^2 = e^{-{d}(n-1)/\chi}\frac{\cos(2\pi\tilde d n)-\cosh({\tilde d n}/{\chi})}{\cos(2\pi \tilde d)-\cosh({\tilde d}/{\chi})},
    \label{eq:analyticVns}
\end{gather} 
where $J_n^*$ denotes the complex conjugate of $J_n$, with $J_n$ representing the complex amplitude of the leading Fourier component of the flapping velocity required to keep a constant separation. If $|J_n|<1$ then flyer $n$ operates more efficiently than in isolation, in the sense that it attains the same forward speed while executing a reduced-amplitude vertical motion. Conversely, $|J_n|>1$ indicates reduced efficiency compared with solo flight.

In the large-separation limit $d\to\infty$, $J_n\to1$ and $V_n^{(0)}(t)\to\cos(2\pi t)$, as expected. In the strong-interaction limit $\chi\to\infty$ the exponentials tend to unity and the sum simplifies to
\begin{align}
    V_n^{(0)}(t)\to \cos\left(\pi \left[2t- \tilde d n+\tilde d\right]\right)\frac{\sin(\pi \tilde d n)}{\sin(\pi \tilde d)},\qquad \chi\to\infty.
\end{align}
Finally, as $n\to\infty$ the leading-order flapping velocities~\eqref{eq:analyticVns} converge to
\begin{align}\label{eq:J_infinity}
    V_\infty^{(0)}(t)=\frac{1}{2}\left( J_\infty e^{2\pi i t}+J_\infty^\ast e^{-2\pi i t}\right),\qquad J_\infty=\frac{1}{1-e^{-\tilde d(1/\chi+2\pi i)}},\qquad\vert J_\infty \vert^2 = \frac{1}{2}\frac{e^{{\tilde d}/{\chi}}}{\cosh({\tilde d}/{\chi})-\cos(2\pi\tilde d)}.
\end{align}
Convergence to this limit is exponential in $n\tilde d/\chi$, so the limiting flapping velocity is expected to emerge in relatively small formations.

\subsection{Flapping out of phase}\label{sec:outphase}

\begin{figure}[htpb!]
    \centering
    \includegraphics[width=\linewidth]{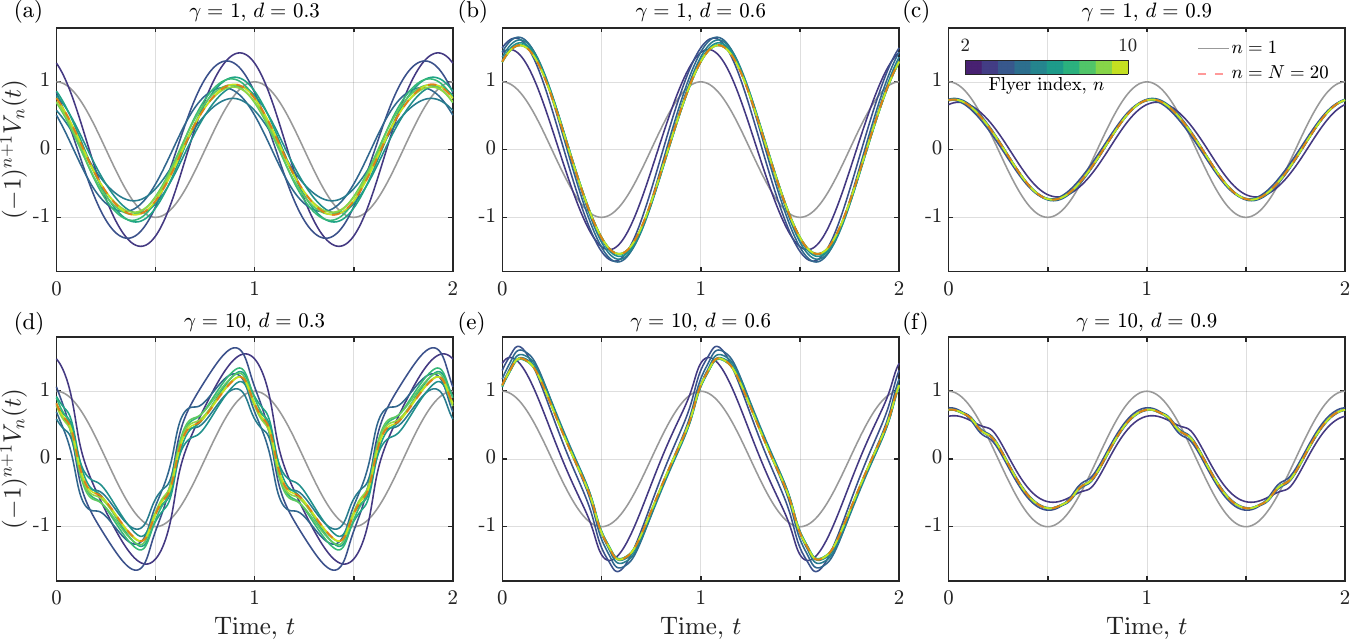}
    \caption{Same as in figure~\ref{fig:Vn_gamma_d_inphase}, but with sign-modified flapping velocities $(-1)^{n+1}V_n(t)$, obtained using \eqref{eq:GeneralFormulaVn}.}
    \label{fig:Vn_gamma_sweep_out_phase}
\end{figure}

For the same sinusoidal flapping velocity for the leader, $V_1(t)=\cos(2\pi t)$, we now analyse the alternative situation for which ${\sigma_n = (-1)^{n+1}}$. In figure~\ref{fig:Vn_gamma_sweep_out_phase} we plot the sign-modified velocities $(-1)^{n+1}V_n(t)$ for two values of $\gamma$ (1 and 10) and three values of $d$ (0.3, 0.6, and 0.9), with $\chi=1$. For small $\gamma$ the velocities are again dominated by the first Fourier mode and resemble cosines, whereas for large $\gamma$ higher-frequency phenomena appear.
However, some important differences may be observed. 

Due to the alternating sign choice, the raw flapping profiles $V_n(t)$ do not converge as $n\to\infty$. However, the sign-modified profiles $(-1)^{n+1}V_n(t)$ do converge, and the corresponding limit is shown as a dashed red curve in figure~\ref{fig:Vn_gamma_sweep_out_phase}. This is why we plot the sign-modified quantity rather than $V_n(t)$ itself.

A striking difference between the in-phase and out-of-phase arrangements is the role of the separation $d$. In the out-of-phase case the amplitude of $V_n(t)$ is large for $d=0.6$ and small for $d=0.9$, the opposite trend to that observed for in-phase flapping. This indicates that the parameter $d$ strongly influences which sign choice is energetically favourable. In particular, one sign choice may incur significantly lower energy expenditure, since energy costs scale roughly with $V_n(t)^2$. Perhaps even more crucially, it might be the case that for a given $d$ one of the two choices for the flapping velocity is \emph{dynamically stable} while the other is \emph{unstable}. By instability we mean that small perturbations in the flyer separation grow and lead to qualitatively different outcomes: (i) collision, where $d_n(t)=0$ at some time; (ii) adjustment to a nearby equilibrium separation; or (iii) loss of cohesion, where $d_n(t)\gg d$. Examples of these scenarios are shown in figure~\ref{fig:numerical_stability}. We develop these ideas further in \S\ref{sec:StabilityMainText}. To understand the role of $d$, we now perform a similar analysis to that in \S\ref{sec:inphase} to evaluate $V_n(t)$ analytically for small and moderate values of $\gamma$.

\begin{figure}[htpb!]
    \centering
    \includegraphics[width=\linewidth]{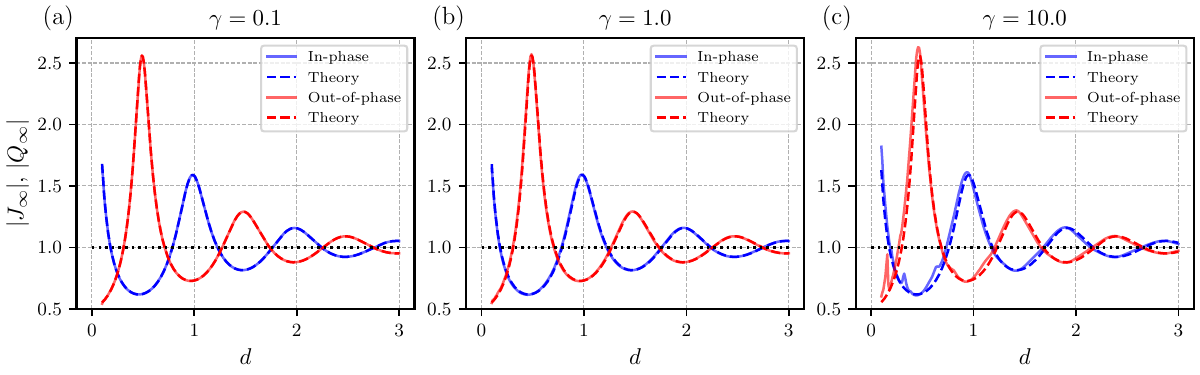}
    \caption{Plots of the first Fourier coefficient of $V_\infty(t)$ for the in-phase flapping velocities ($\sigma_n=+1$; blue) and the out-of-phase flapping velocities ($\sigma_n=(-1)^{n+1}$; red) versus fixed flyer separation $d$. Analytical approximations to each one, expected to hold for $\gamma\lesssim 1$ are plotted as dashed lines. Results are shown for $\chi=1$ and for $\gamma$ equal to (a) 0.1, (b) 1, and (c) 10.}
    \label{fig:in_phase_out_phase_comparison}
\end{figure}

Proceeding as in \S\ref{sec:inphase}, we approximate $t_2(t)=t-\tilde d+O(R)$ and expand $V_n(t)=V_n^{(0)}(t)+R\,V_n^{(1)}(t)+\dots$. The leading-order term becomes
\begin{align}
    V_n^{(0)}=\sum_{j=0}^{n-1}(-1)^{n-j+1}\cos\left(2\pi\left[t-j\tilde d\right]\right)e^{-\tilde d j/\chi},
\end{align}
which may be summed to give
\begin{align}
    V_n^{(0)}(t)=\frac{1}{2}\big(Q_n e^{2\pi i t}+Q_n^*e^{-2\pi i t}\big),\qquad Q_n=(-1)^{n+1}\frac{1+(-1)^{n+1}e^{-\tilde d\, n(1/\chi+2\pi i)}}{1+e^{-\tilde d(1/\chi+2\pi i)}}.
\end{align}
In the strong-interaction limit with $\chi\to\infty$ this simplifies to
\begin{align}
    V_n^{(0)}(t)=(-1)^n\cos\left(\pi\left[2t-\frac{n}{2}-\tilde d (n-1)-\frac{1}{2}\right]\right)\frac{\sin\left(\pi n [d+1/2]\right)}{\cos(\pi \tilde d)}.
\end{align}
Finally, taking $n\to\infty$ yields
\begin{align}\label{eq:Q_infinity}
  (-1)^{n+1}Q_n \to\frac{1}{1+e^{-\tilde d(1/\chi+2\pi i)}},\qquad \vert Q_\infty\vert^2=\frac{1}{2}\frac{e^{\tilde d/\chi}}{\cosh(\tilde d/\chi)+\cos(2\pi \tilde d)}.
\end{align}
Figure~\ref{fig:in_phase_out_phase_comparison} compares the first Fourier coefficient of $V_\infty(t)$ for the in-phase and out-of-phase arrangements as a function of the prescribed separation $d$, and contrasts numerical results with the approximations $|J_\infty|$ from~\eqref{eq:J_infinity} and $|Q_\infty|$ from \eqref{eq:Q_infinity}. For $\chi\gtrsim1$, the in-phase amplitude is smaller for approximately $d\in(0.25,0.75)\cup(1.25,1.75)\cup\cdots$, while the out-of-phase amplitude is smaller for $d\in(0,0.25)\cup(0.75,1.25)\cup\cdots$. In those ranges the amplitudes are less than unity, meaning the formation is energetically advantageous compared with flying solo. Conversely, where a strategy yields amplitude exceeding unity it would be energetically preferable to fly alone. Because the period-averaged energy usage increases with flapping amplitude, the separation $d$ plays a key role in determining which flapping strategy is more efficient. 

Therefore, at first glance, the energetically preferred choice for a given $d$ is the strategy with the smaller amplitude. However, as we demonstrate below, the energetically favourable choice can be dynamically unstable, and stability must therefore be taken into account when deciding which strategy is viable for sustained formation flight.

\subsection{In or out of phase? Stability of solutions determines the correct choice}\label{sec:StabilityMainText}
Each follower has two possible sign choices $\sigma_n=\pm1$ that enforce the constant-separation constraint with the upstream neighbour. We studied the in-phase choice $\sigma_n=+1$ and the out-of-phase choice $\sigma_n=(-1)^{n+1}$ and found that, for a given separation $d$, one of these typically requires a smaller flapping amplitude. This suggests an energy argument for selecting a strategy. In this subsection we show that stability considerations can supersede the energy argument.

Specifically, we will consider whether each of the two flapping strategies are stable to small perturbations in the flyer separation. For analytical tractability we study the two-flyer problem. The leader flaps according to $V_1(t)=\cos(2\pi t)$ and the follower is prescribed the separation-preserving flapping
\begin{align}
    V_2(t)=\sigma_2 V_1(t)+V_1\bigl(X_1^{-1}(X_1(t)-d)\bigr),
\end{align}
with $\sigma_2=\pm1$. We perturb the initial position of the follower by a small amount,
\begin{align}
    X_2(0)=-d+\epsilon, \qquad U_2(0)=U_1(0),\qquad \text{ with }\vert\epsilon \vert\ll1,
\end{align}
and define the configuration as stable if $X_1(t)-X_2(t)\to d$ as $t\to\infty$, and unstable if the perturbation grows. The small perturbation $\epsilon$ may be positive or negative.

Physically, this corresponds to asking whether a follower that \emph{believes} it is at distance $d$ but is actually at $d+\epsilon$ will choose a flapping velocity that amplifies the error. This is a natural question because sensory and measurement errors are unavoidable in biological systems, and any practical separation-preserving rule must be robust to such errors. Similar statements can be made for engineering situations, such as drone formations during lighting displays \cite{lanteigne2017design,zhihao2020virtual,ioc_drone_show_2021,ars_drone100_2016}.

\begin{figure}[htpb!]
    \centering
    \includegraphics[width=\linewidth]{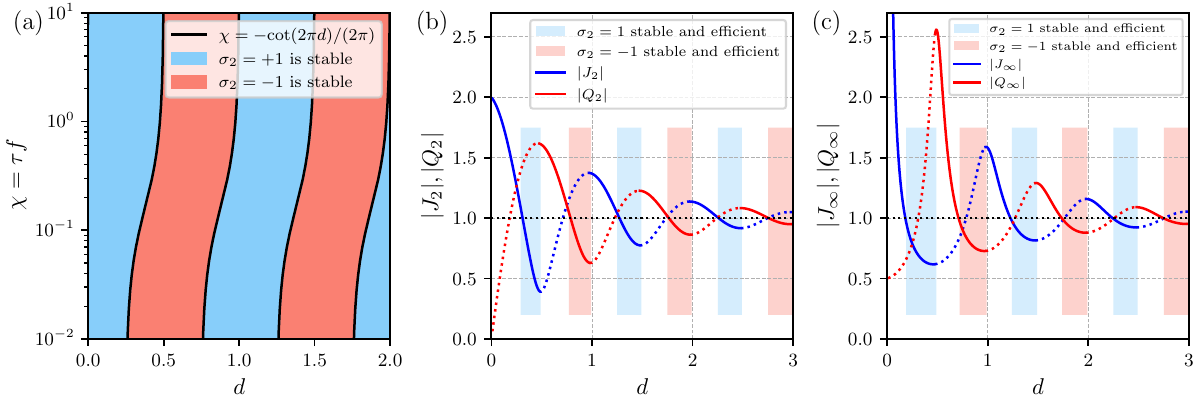}
    \caption{(a) Stability diagram in $(d,\chi)$ space. We plot whether the sign of $S/\sigma_2$  as given in \eqref{eq:signofS_stability} is positive (blue, in-phase is stable) or negative (red, out-of-phase is stable). (b) Magnitude of flapping velocities for in-phase ($J_2$) and out-of-phase ($Q_2$) flapping as a function of $d$, with ``goldilocks'' regions of both stability and efficiency highlighted in light blue and light red shades. Lines are drawn solid where they are stable and dotted where they are unstable. (c) Same as panel (b) but with $J_\infty$ and $Q_\infty$ instead. In panels (b,c), $\chi=1$.}
    \label{fig:StabilityDiagram}
\end{figure}

To study the stability of the formation to perturbations in the initial conditions, we use a combined numerical and analytical approach. For the theory, appendix~\ref{sec:AppStability} contains an asymptotic expansion in powers of $\epsilon$ that yields linearised equations for the perturbations. Focusing on $\gamma\lesssim1$, where $t_2(t)=t-\tilde d+O(R)$, we derive a simple stability criterion: the linearised dynamics are unstable when
\begin{align}\label{eq:signofS_stability}
    S(\sigma_2,d,\chi)=-\sigma_2\big[\cos(2\pi d)+2\pi \chi \sin(2\pi d)\big]<0.
\end{align}
An important conclusion follows immediately: only one of the two sign choices is stable for any given $(d,\chi)$. The stability boundary is given by $\chi(d)=-\cot(2\pi d)/(2\pi)$, and crossing this curve switches which sign is stable. Figure~\ref{fig:StabilityDiagram}(a) maps the regions in $(d,\chi)$ space where in-phase flapping ($\sigma_2=+1$) or out-of-phase flapping ($\sigma_2=-1$) is stable. In appendix~\ref{sec:AppStability} we confirm these predictions by solving the full nonlinear perturbed problem~\eqref{eq:StabilityEquationFull}; the agreement for the values we tested is excellent.

Most interestingly, the stable strategy does not always coincide with the energetically optimal strategy. Figures~\ref{fig:StabilityDiagram}(b) and \ref{fig:StabilityDiagram}(c) show parameter ranges where the stable solution is also the most efficient, i.e. $|J|<1$ or $|Q|<1$. We refer to these ranges as ``\emph{goldilocks}'' zones and speculate that natural flyers might operate preferentially in such ranges.

An additional observation is that the most efficient amplitudes (the troughs in the blue and red curves) lie close to the stability boundary, typically at the right edge of the goldilocks zones. This implies that operating near peak efficiency is associated with an elevated risk that a small measurement error will push the system into an unstable regime. Consequently, natural or engineered formations may trade off some efficiency for a stability margin by selecting separations to be near the centre of the goldilocks zones.

\subsection{Connection to stable emergent equilibrium distances}\label{sec:connectionEqmDist}

The primary focus of previous studies has been on flying formations in which the flyers have a prescribed flapping motion given by $V_n(t)=\cos(2\pi t)$ and the propulsive speed $U_n(t)$ and $d_n(t)=X_{n-1}(t)-X_n(t)$, the gap distance between flyer $n$ and its upstream neighbour $n-1$ (shown schematically in figure~\ref{fig:modelSchem}(c)), emerge as a consequence of the flow interactions among group members. 

Following the same process as in our previous work~\cite{mavroyiakoumou2025modeling}, we derive here the equilibrium spacing between two flyers if we assume that the flapping motion for both is  $V_1(t)=V_2(t)=\cos(2\pi t)$. The thrust per unit mass for the follower $(n=2)$ is given by 
\begin{align}
    B_2&=2\gamma\left(V_2(t)-V_1(t-\Delta t)e^{-\Delta t/\chi} \right)^2\nonumber\\
    &=2\gamma\left(\frac{1}{2}+\frac{1}{2}e^{-2\Delta t/\chi}-\cos(2\pi \Delta t)e^{-\Delta t/\chi} \right.\nonumber\\
    &\left.\phantom{=(((2\gamma }+ \frac{1}{2}\cos(4\pi t)+\frac{1}{2}\cos(4\pi [t-\Delta t])e^{-2\Delta t/\chi}-\cos\left(4\pi \left[t-\frac{1}{2}\Delta t \right]\right)e^{-\Delta t/\chi}\right).\label{eq:B2stableDist}
\end{align}
If we neglect the time-dependent terms in~\eqref{eq:B2stableDist}, the time-independent terms correspond to: (a) the self-generated thrust from the flapping, which is $\gamma$, and (b) the thrust generated as a result of the wake interactions, given by $\gamma(e^{-2\Delta t/\chi}-2\cos(2\pi \Delta t)e^{-\Delta t/\chi})$. At equilibrium, we have that $U^*_2=U^*_1$, which requires that the interaction thrust is zero.

To compute steady equilibrium positions $d^*$ when the flyers are flapping in phase, we find solutions to the transcendental condition~\cite{mavroyiakoumou2025modeling}:
\begin{equation}\label{eq:transcendentalS}
    \cos(2\pi d^*)=\frac{1}{2}e^{ -d^*/\chi},
\end{equation}
where we use that $\Delta t^*=d^*$. We may obtain an identical equation \eqref{eq:transcendentalS} for the steady equilibrium positions by imposing $\vert J_\infty\vert=1$, with $\vert J_\infty\vert$ given in \eqref{eq:J_infinity}. We could therefore also find these values of $d^*$ for $\chi=1$ by identifying the points where the blue line intersects $y=1$ in figure~\ref{fig:in_phase_out_phase_comparison}.

For out-of-phase flapping, a similar calculation gives the corresponding condition
\begin{equation}\label{eq:transcendentalSout}
    \cos(2\pi d^*)=-\frac{1}{2}e^{ -d^*/\chi},
\end{equation}
which can also be obtained by imposing $|Q_\infty|=1$ with $Q_\infty$ defined in \eqref{eq:Q_infinity}. Thus, the equilibrium separations obtained from the emergent spacing problem are directly related to the separation-preserving amplitudes: points at which $|J_\infty|=1$ or $|Q_\infty|=1$ correspond to neutral interaction thrust and hence to candidate equilibrium spacings.

\begin{figure}[htpb!]
    \centering
     \includegraphics[width=\linewidth]{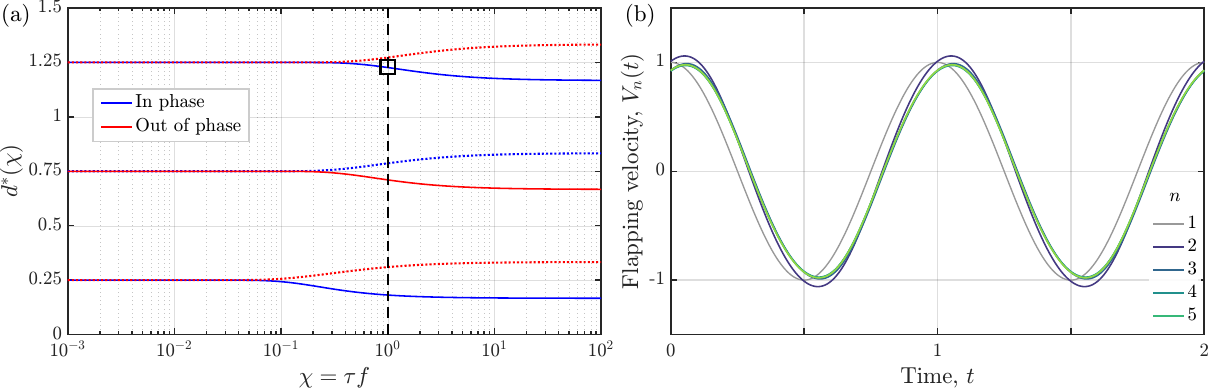}
    \caption{(a) Solution branches $d^*(\chi)$ to the transcendental condition \eqref{eq:transcendentalS} for a wide range of values of $\chi$. (b) Enforcing as the gap distance $d$ the analytically predicted stable dimensionless gaps $d^*$, we find that the flapping velocities $V_n(t)$ for all flyers $n=1,2,\dots,5$ are almost the same. Here, we use $\gamma =1.315$, $\chi=\tau f=1$, $d^*=1.227$ (with $\chi$ and $d^*(\chi)$ corresponding to the black square in panel (a)).}
    \label{fig:EmergentEquilibriumDistances}
\end{figure}

Figure~\ref{fig:EmergentEquilibriumDistances}(a) shows solution branches $d^*(\chi)$ of the transcendental conditions \eqref{eq:transcendentalS} and \eqref{eq:transcendentalSout} for a range of $\chi$, with in-phase branches shown in blue and out-of-phase branches in red. We indicate stability using solid lines for stable branches and dotted lines for unstable ones, based on the stability analysis in \S\ref{sec:StabilityMainText}.

In figure~\ref{fig:EmergentEquilibriumDistances}(b) we enforce one of the analytically predicted stable dimensionless gaps $d^*\approx 1.227$ (for ${\chi=1}$) and plot the resulting $V_n(t;d^*)$. We observe that, while $V_n(t)$ for large $n$ closely matches the leader's sinusoidal flapping velocity in amplitude, a small phase shift is present for the first few followers. This effect can be attributed to leader-edge effects in a linear array and to the approximations made when deriving the transcendental conditions, which neglect certain time-dependent terms.

\section{Non-band-limited leader flapping}\label{sec:higherModes}
So far, we have focused on leader flapping velocities that consist of a single oscillatory frequency. In practice, however, a flyer's vertical motion need not be purely sinusoidal and may contain multiple harmonics. We therefore consider the more general case in which the leader flaps with an arbitrary smooth and periodic velocity that is not band-limited.

Assuming periodicity with unit period, the leader's flapping velocity may be expressed as a Fourier series
\begin{align}\label{eq:LeaderFlappingCk}
    V_1(t)=
    \sum_{k\neq0}c_k e^{2\pi i k t},\qquad c_{-k}=c_k^\ast,
\end{align}
where the zero mode is set to zero, $c_0=0$, so that the vertical motion has zero mean. To remain consistent with the non-dimensionalisation adopted earlier, we impose the normalisation
\begin{align}
    \int_0^1 [V_1(t)]^2 \,\upd t= \sum_{k=-\infty}^{\infty} |c_k|^2= \frac{1}{2},
\end{align}
where the first equality follows from Parseval's theorem \cite{stein_shakarchi_fourier}. This is a normalisation constraint used to ensure the period-averaged propulsive velocity is (nearly) unity, consistent with our choice of scalings.

The recursive formula \eqref{eq:RecursiveFormulainTermsoft_2}, or equivalently the explicit expression \eqref{eq:GeneralFormulaVn}, determines the separation-preserving flapping velocities of the followers. In this section we analyse how the spectral content of $V_1(t)$ propagates down through the formation and how it depends on the separation distance $d$ and the parameter $\gamma$.

\subsection{Moderate $\gamma$ values}
We first consider the regime of small to moderate $\gamma$, specifically $\gamma\lesssim 1$. In this case, the leader's propulsive velocity is nearly constant, $U_1(t)\approx 1$, and therefore $t_2(t)\approx t-d.$
Substituting this approximation into the general expression \eqref{eq:GeneralFormulaVn} with the in-phase choice $\sigma_n=+1$, we obtain
\begin{align}
    V_n(t)=\sum_{j=0}^{n-1}e^{-jd/\chi}\sum_{k\neq0}c_k e^{2\pi i k(t-jd)}.
\end{align}
Reordering the sums and evaluating the resulting geometric series yields
\begin{align}\label{eq:VnModerateGammaCk}
    V_n(t)=\sum_{k\neq 0}c_k\sum_{j=0}^{n-1}e^{-dj/\chi}e^{2\pi i k(t-jd)}=\sum_{k\neq 0}c_ke^{2\pi i kt}\sum_{j=0}^{n-1}e^{-dj/\chi}e^{-2\pi i kjd}=\sum_{k\neq 0}c_ke^{2\pi i kt}\left[\frac{1-e^{-dn\left(1/\chi+2\pi i k\right)}}{1-e^{-d\left(1/\chi+2\pi i k\right)}}\right],
\end{align}
Equation~\eqref{eq:VnModerateGammaCk} shows that each Fourier mode of the leader flapping velocity is preserved in the followers, but with a separation-dependent amplitude modification. Specifically, the complex Fourier coefficients transform according to
\begin{align}\label{eq:CkCoefficientsVn}
    c_k\longmapsto c_k\left[\frac{1-e^{-dn\left(1/\chi+2\pi i k\right)}}{1-e^{-d\left(1/\chi+2\pi i k\right)}}\right].
\end{align}

Since this factor does not decay with $k$, higher harmonics present in the leader's motion persist throughout the formation. Thus, non-band-limited flapping by the leader propagates downstream without spectral filtering. Moreover, the appearance of factors involving $e^{2\pi i k d}$ indicates that the separation distance $d$ plays a crucial role in selectively amplifying or attenuating individual harmonics.

\subsection{Two frequencies}
To illustrate the influence of higher harmonics, we consider a simple example in which the leader flaps with two frequencies of equal weight,
\begin{align}
\label{eq:leaderharmonics}
    V_1(t)=\frac{1}{\sqrt{2}}\bigl(\cos(2\pi t)+\cos(4\pi t)\bigr).
\end{align}
corresponding to complex Fourier coefficients $c_1=c_2=1/(2\sqrt{2})$.

\begin{figure}[t]
    \centering
    \includegraphics[width=\linewidth]{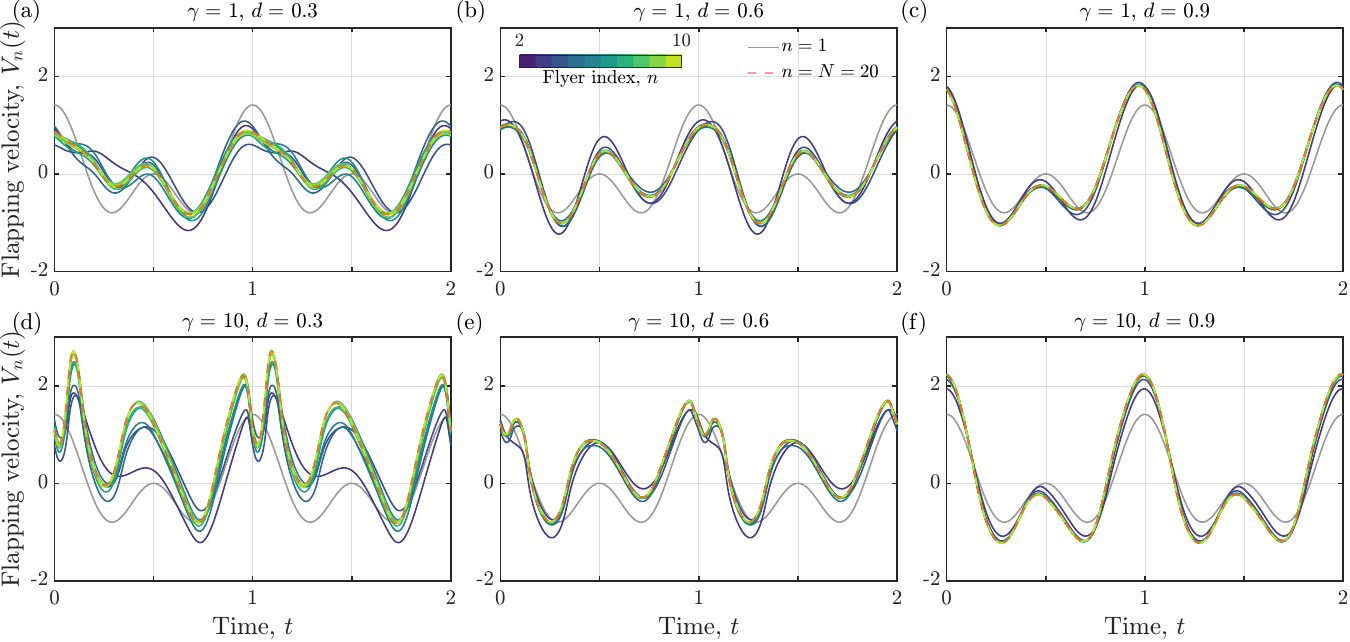}
    \caption{Flapping velocities $V_n(t)$ for flyers $n\geq2$ required to maintain a constant separation $d$ from their upstream neighbour when the leader flaps with a two-frequency flapping velocity given by $V_1(t)$ in \eqref{eq:leaderharmonics} (grey curves). The limiting downstream profile $V_\infty(t)$ is shown as a dashed red curve. Parameter values: $\chi=1,\,\sigma_n=+1$.}
    \label{fig:VnHighHarmonics}
\end{figure}

We show in figure \ref{fig:VnHighHarmonics} the flapping velocities $V_n(t;\gamma,d)$ for $n\geq 2$, when the first flyer flaps with velocity given by \eqref{eq:leaderharmonics} (plotted in grey) and $V_n(t)$ is obtained recursively through \eqref{eq:RecursiveFormulainTermsoft_2} with $\sigma_n=+1$. 
Compared to the single-frequency case, we observe that substantially richer temporal structure develops in the followers' flapping velocities, particularly for larger values of $\gamma$ (figures~\ref{fig:VnHighHarmonics}(d) and~\ref{fig:VnHighHarmonics}(e)). As $\gamma$ increases, nonlinear effects in the propulsion dynamics introduce additional higher harmonics that are not present in the leader's motion. This indicates that maintaining controlled formation flight becomes increasingly complex when the leader's flapping is spectrally broad.

\begin{figure}[t]
    \centering
    \includegraphics[width=\linewidth]{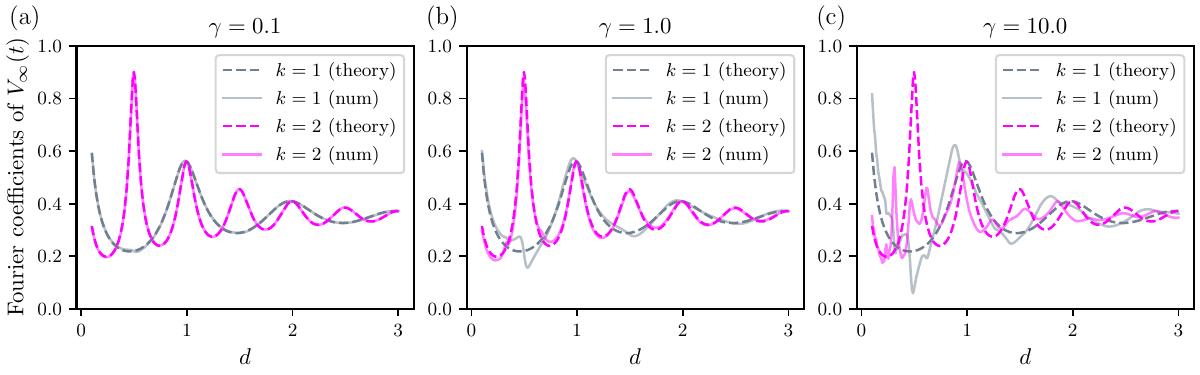}
    \caption{Magnitude of the first two Fourier coefficients of the limiting flapping velocity $V_\infty(t)$ as functions of the separation distance $d$ for $\gamma$ equal to (a) $0.1$, (b) $1$, and (c) $10$, with $\chi=1$. Solid curves show numerical results obtained from the full nonlinear model, while dashed curves show the theoretical predictions from \eqref{eq:CkCoefficientsVn}, valid for $\gamma\lesssim1$.}
    \label{fig:CkFunctionofD}
\end{figure}

To quantify this effect, we compute $V_\infty(t;d,\gamma)$ numerically for $\gamma=0.1,1,$ and $10$, and for separation distances $d\in(0,3)$. In particular, we plot in figure~\ref{fig:CkFunctionofD} the magnitudes of the first two Fourier coefficients of $V_\infty(t)$ as functions of $d$ (solid lines), and compare them with the theoretical predictions from \eqref{eq:CkCoefficientsVn} (dashed lines).

For $\gamma\lesssim 1$ (figures~\ref{fig:CkFunctionofD}(a) and~\ref{fig:CkFunctionofD}(b)), the theoretical expressions accurately capture the dependence of the Fourier amplitudes on the separation distance, with only minor deviations near resonant values of $d$. In contrast, for $\gamma\gg 1$ (figure~\ref{fig:CkFunctionofD}(c)) the agreement deteriorates, as nonlinearities generate additional frequency content beyond that predicted by the linearised analysis. This highlights the increasing importance of nonlinear effects in strongly forced regimes and suggests that separation-preserving strategies become more sensitive to the leader's kinematic details when $\gamma$ is large.

\section{Discussion and conclusions}\label{sec:conclusions}

We have studied the dynamical consequences of wake interactions in the case of oscillating propulsors arranged in linear arrays, where the flyers flap in such a way as to maintain a constant separation distance between them. Configurations in which multiple flyers flap in phase with identical sinusoidal kinematics are intrinsically fragile. In such systems, many-body flow interactions give rise to flonons, which are self-amplifying collective waves, that can trigger global destabilisation across the array~\cite{newbolt2024flow}. Specifically, previous work has shown that the physical interactions are important in setting the overall structure of flying formations, but that sensing-and-response is required to maintain or stabilise them~\cite{nitsche2025stability,mavroyiakoumou2025modeling}. Therefore, in this work, with the flyers freely choosing their flapping kinematics in order to keep a constant separation distance between them, it is possible to obtain arbitrarily long arrays.

Starting from the interaction model introduced in our previous work \cite{mavroyiakoumou2025modeling}, we derived an asymptotic approximation for an isolated flyer that is valid for $\gamma\sim1$. Using this approximation to avoid large transients, we then determined the separation-enforcing flapping velocities exactly by imposing equality of forward, propulsive velocities between consecutive flyers. This was made possible by reformulating the follower-wake model as a system of ordinary differential equations solved iteratively. Each follower has a binary sign choice for its vertical flapping velocity, so that there are two distinct $V_n(t)$ functions that enforce constant separation with an upstream neighbour. For a sinusoidally flapping leader, we summarise the main behaviours we observed:
\begin{enumerate}[label=(\roman*)]
\itemsep0em
    \item For $\gamma\lesssim 1$, the follower profiles $V_n(t)$ remain close to sinusoidal, and their oscillation amplitudes are strongly determined by the separation $d$ and by the sign choice.
    \item For $\gamma\gg 1$, the follower profiles become intricate periodic functions that contain multiple fundamental frequencies; even harmonics are generally negligible and odd harmonics dominate.
\end{enumerate}

The sign choice, together with the separation distance, strongly affect the required flapping amplitude and, therefore, the energetic cost of group flight. When the dimensionless flapping amplitude is less than unity, group flight is energetically favourable compared to isolated flight; when it exceeds unity, flying in formation becomes less efficient. At first glance this suggests an energy-based rule for selecting the sign. The main finding of this work is that energy considerations alone are insufficient because, for any given parameter set $(d,\chi)$ and for $\gamma\lesssim1$, only one of the two sign choices is dynamically stable. In the two-flyer linearised problem we obtain the simple criterion
\begin{align}
    -\sigma_2\bigl[\cos(2\pi d)+2\pi\chi\sin(2\pi d)\bigr]>0,
\end{align}
which selects the stable sign $\sigma_2=\pm1$. The stable sign does not always coincide with the energetically optimal sign, so there are regions of parameter space where group flight is not viable either because of energetic cost or because of lack of stability. We identify ranges of separation in which the stable sign is also energy efficient (termed goldilocks zones). It is plausible that natural flyers have evolved to operate within goldilocks zones where stability and aerodynamic benefit are compatible.

We also examined the case in which the leader flaps with a spectrally broad periodic profile. When the leader contains multiple frequencies, the separation-preserving profiles become substantially more complicated. Nonlinear interactions and aliasing effects amplify higher harmonics and lead to complex solutions even for moderate $\gamma\sim1$ when the separation is small ($d<0.5$). Maintaining formation flight under such forcing is therefore significantly more challenging than under purely sinusoidal forcing. 

An important practical question concerns the sensory requirements of a follower that must maintain a prescribed separation. First, the flyer must estimate the ambient wake velocity produced by its upstream neighbour, and this requires appropriate mechanosensory capability. Fluid--structure sensors typically sense velocity only cleanly within specific parameter ranges, so the effective design of such sensors and their mechanical tuning matter for reliable operation \cite{emde_senses_2012,chico-vazquez_uncovering_2025}. Second, after sensing the wake, the follower must select the sign choice $\sigma_n=\pm1$ that yields dynamical stability. Because stability depends strongly on separation $d$, the flyer must also estimate this distance. In biological systems this may be achieved visually or by other sensory modalities.

There are several new avenues for future work. The trade-off between efficiency and instability risk merits systematic optimisation. A flyer may prefer a slightly suboptimal energetic choice if that increases the stability margin and reduces the risk of collision. This trade-off could be studied by combining the present stability criteria with explicit energetic measures, similarly to recent work that balances energy expenditure and crash risk in drafting-mediated motion \cite{chico-vazquez_mathematical_2025}. Additionally, our analysis focused on linear arrays with a distinguished leader, but another important configuration to consider, mostly relevant to flapping foil experiments~\cite{becker2015hydrodynamic,ramananarivo2016flow,newbolt2024flow}, is a periodic, closed chain, in which leader-edge effects are absent and every flyer is dynamically equivalent. The separation-preserving rules in that setting may admit additional symmetric solutions and different stability properties. Our results could be extended to include more general drag laws of the form $D\sim U^\beta$, to model a broader range of physical systems. We expect results there to be qualitatively similar, with differing quantitative constants (and the crucial dimensionless parameter $\gamma$ will also be different).

The model we employed is deliberately simplified to capture the essential role of wake memory and nearest-neighbour, downstream forcing, but these simplifications impose limitations that should be borne in mind. First, the aerodynamic model simplifies wakes as decaying one-dimensional signals. Real wakes are three dimensional and can interact in more complex, non-local ways. Second, the drag law and thrust scaling used here were chosen to reproduce laboratory flapping-foil experiments, and other regimes of Reynolds number or different vehicle geometries may lead to quantitatively different dimensionless parameters. Third, we assumed that followers can instantaneously implement the prescribed flapping profile once they choose a sign; in practice actuators have finite bandwidth and sensorimotor delays that can alter stability boundaries. Fourth, our principal stability results are derived in the small-perturbation, small-$\gamma$ asymptotic regime and are therefore local in nature; large perturbations, heterogeneous fleets with parameter variability (i.e.\ different flyer mass), or stochastic forcing may yield different outcomes. Finally, we considered only nearest-neighbour one-way interactions; two-way coupling and longer-range hydrodynamic interactions could qualitatively modify both the energetics and the stability. Addressing these limitations will require higher-fidelity fluid models, experimental tests on robotic flapping platforms, and stochastic or robust-control extensions of the present theory.

In summary, our work clarifies how simple separation-preserving rules translate into nontrivial collective kinematics and how the interplay between energetic benefit and dynamical stability determines viable formation strategies. The results inform both biological hypotheses about how animals might tune spacing and kinematics in groups and engineering design for teams of flapping-wing vehicles/robots that must fly in close formation.

\appendix
\section{Table with typical values}

For the model results reported in the main text, we have employed parameter values that fall in the ranges estimated from flapping-foil experiments in water~\cite{becker2015hydrodynamic,ramananarivo2016flow,newbolt2019flow,newbolt2024flow} and summarised in Table~\ref{tab:ParameterValues}. Using these, we also estimate the values of $\gamma$ and $\chi$, which are the main dimensionless parameters dictating the flyer dynamics in our current work.

\renewcommand*{\arraystretch}{1.05}
\setlength{\extrarowheight}{1pt}
\setlength{\LTcapwidth}{\textwidth}
\begin{longtable}{lll}
\caption{Dimensional parameters and variables, and dimensionless groups used in the follower-wake interaction model. The symbol ``E'' in the third column is used to denote values that were used in previous experiments of flapping foils in a water tank~\cite{becker2015hydrodynamic,ramananarivo2016flow,newbolt2019flow,newbolt2024flow}.}
\vspace*{.3cm} \\
\hline
\textbf{Symbol} & \textbf{Definition/meaning} & \textbf{Typical value/range} \\
\hline
\endfirsthead
\multicolumn{3}{l}{\textbf{Table \thetable\ (continued)}}\\
\hline
\textbf{Symbol} & \textbf{Definition/meaning} & \textbf{Typical value/range} \\
\hline
\endhead
\hline
\endfoot
\multicolumn{3}{l}{\textbf{Dimensional parameters}}\\
\hline
$\rho$ & Fluid density &  E: 1 g/cm$^3$\\
$\mu$ & Fluid dynamic viscosity & E: 0.01 g/(cm s) \\
$\nu$ & Fluid kinematic viscosity, $\mu/\rho$ & E: 0.01 cm$^2$/s\\
$M$ & Mass of flyer & 20--80 g, E: 20 g\\
$c$ & Flyer chord length & 3--6 cm, E: 4 cm\\
$s$ & Flyer span & 5--15 cm, E: 8 cm\\
$A$ & Peak-to-peak flapping amplitude & 1--10 cm, E: 3 cm\\
$f$ & Flapping frequency & 0.1--20 Hz, E: 2.5 Hz\\
$U_n(t)$ & Flyer propulsive (forward) speed  $n$ & 15--30 cm/s \\
$V_n(t)$ & Flyer flapping (vertical) speed & \\
$W_n(x,t)$\hspace*{.5cm} & Wake flow velocity generated by flyer $n$ & \\
$t_n(t)$ & Earlier time when wake of flyer $n\!-\!1$ was generated at $X_n$ & \\
$X_n(t)$ & Horizontal position of flyer $n$ & \\
$d_n(t)$ & Gap between flyer $n$ and its upstream neighbour & \\
$\tau$ & Wake decay timescale & E: $0.5\,\mathrm{s}$ \\
$T_n$ & Thrust force on flyer $n$ & \\
$D_n$ & Aerodynamic drag force on flyer $n$ & \\

\hline
\multicolumn{3}{l}{\textbf{Dimensionless groups}}\\
\hline
$C_T$ & Dimensionless thrust coefficient & $0.8$–$1.1$, E: 1 \\
$C_D$ & Dimensionless skin-friction drag coefficient & E: 10 \\
$\chi$ & Ratio of wake dissipation rate to flapping period, $\tau f$ & $>0$, E: 1.25\\
$M^*$ & Dimensionless mass ratio, $M/(\rho c^2 s)$ & 0.05--30, E: 0.16\\
$\mathrm{Re}_f$ & Flapping Reynolds number, $\rho A f c/\mu$ & $10^2$--$10^5$, E: 9430\\
$\mathrm{Wo}$ & Womersley number, $c\left({2\pi f}/{\nu}\right)^{1/2}$ & E: 160\\
$\gamma $& ${(C_D^2C_T\pi^5/2)^{1/3}\mathrm{Re}_f^{2/3}}/({\mathrm{Wo}^{2}M^*})$ & $>0$, E: 1.3
\vspace*{4pt}
\label{tab:ParameterValues}
\end{longtable}

\section{Details for single flyer}\label{sec:AppSingleFlyer}

\begin{figure}[htpb!]
    \centering
    \includegraphics[width=\linewidth]{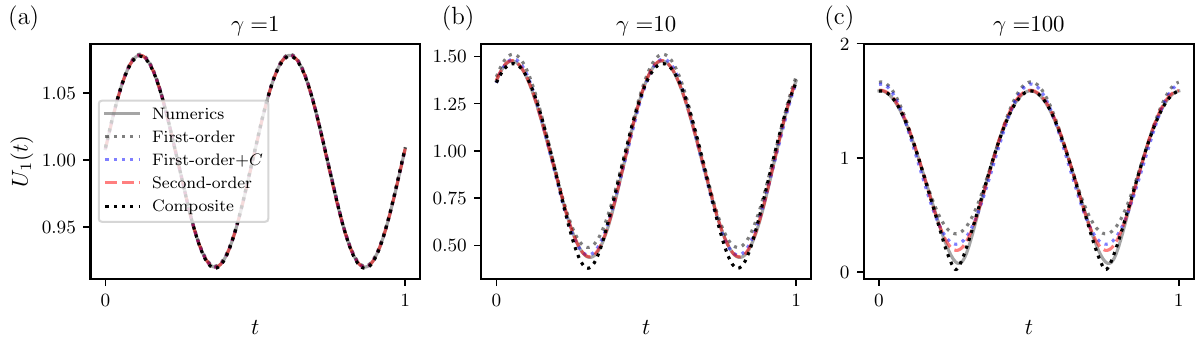}
    \caption{The numerical solution for $U_1(t;\gamma)$, shown as a solid grey line, for three values of $\gamma$ equal to (a) 1, (b) 10, and (c) 100, is compared against increasingly more accurate asymptotic approximations.}
    \label{fig:U1velocities}
\end{figure}

We numerically solve for the propulsive speed of a single flyer $U_1(t)$ using \texttt{solve\_ivp} with method \texttt{DOP853}, absolute tolerance set to $10^{-15}$ and relative tolerance  equal to $5\cdot 10^{-14}$. In figure~\ref{fig:U1velocities}, we plot $U_1(t)$ for $\gamma$ equal to (a) 1, (b) 10 and (c) 100, and compare it with four different approximations:
\begin{enumerate}
\itemsep0em
    \item The first-order approximation given in  \eqref{eq:SingleFlyerFirstOrderVelocity} (dotted grey line).
    \item The first-order approximation with a second-order correction \emph{only} to the period-averaged velocity (and lacking any other second-order terms; dotted blue line).
    \item The full second-order solution (dashed red line).
    \item The composite solution with the correct asymptotic behaviour for $\gamma\ll1$ and $\gamma\gg1$, given in \eqref{eq:CombinedApproximation}  (dotted black line).
\end{enumerate}

\begin{figure}[htpb!]
    \centering
    \includegraphics[width=\linewidth]{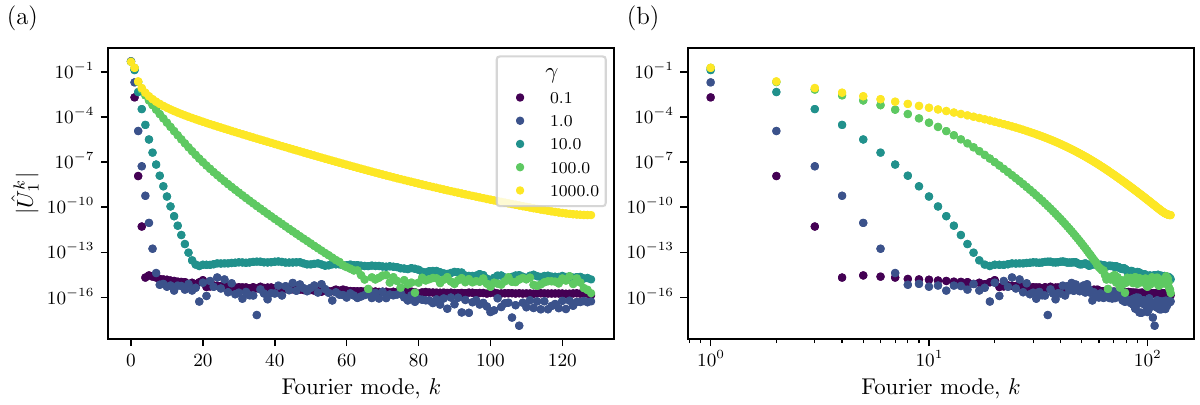}
    \caption{Fourier coefficients of $U_1(t)$ for different values of $\gamma$. The first Fourier coefficient dominates for $\gamma\lesssim 1$. Both panels display the same data: (a) logarithmic $y$-axis, (b) log-log plot (zeroth mode ($k=0$) is not shown). }
    \label{fig:U1FFT}
\end{figure}

We see that for $\gamma\lesssim1$ (figure~\ref{fig:U1velocities}(a)) all approximations are valid, and in particular the first-order approximation is as good as any other approximation. For $\gamma=10$ (figure~\ref{fig:U1velocities}(b)), the large-$\gamma$ approximation (dotted black line) overestimates the amplitude of the solution, and the full second-order solution (dashed red line) offers little improvement over the first-order solution with period-averaged velocity modified by $C$, with $C$ given in \eqref{eq:A0BandC}, but both of these perform better than the exclusively first-order solution. Finally, for very large $\gamma$ values such as $\gamma=100$ (figure~\ref{fig:U1velocities}(c)), only the composite solution that incorporates the nonlinear behaviour near $U_1=0$ can capture the solution. We conclude that for the regime of interest ($\gamma\approx1$) the first-order solution with the period-averaged velocity correction offers a good compromise between accuracy and simplicity.
A second way of ensuring our asymptotic solution is valid is to study the fundamental frequencies of $U_1(t)$. In figure~\ref{fig:U1FFT} we show the magnitude of the Fourier coefficients of $U_1(t)$ for $\gamma=10^j$, $j=-1,\dots,3$, both in a logarithmic $y$-axis (panel (a)) and as a log-log plot (panel (b)). In particular, expanding
\begin{align}
    U_1(t;\gamma)=\sum_{k=-\infty}^\infty \hat U_1^k(\gamma) e^{4\pi i k t},
\end{align}
we plot $\hat U_1^k(\gamma)$ against $k$. We see that for small and moderate values of $\gamma$ there is a sharp exponential decay towards machine precision after less than 10 modes, suggesting indeed that only the first few frequencies are relevant, as predicted from our asymptotic analysis. For large values of $\gamma$ the picture is very different. As we expect $U_1\sim (1+\cos(4\pi t))^{2/3}$, the spectrum is broad and the decay of the $\hat U_k$ is very slow, due to the singular behaviour when $U_1$ approaches zero. In figure~\ref{fig:U1FFT}(a) we see the coefficients converge exponentially for large $k$, and in figure~\ref{fig:U1FFT}(b) we highlight the size of the first few coefficients, showing that the dynamics are dominated by $k=1$ for $\gamma\lesssim 1$.

\section{Details on the computation of $t_2(t)$}\label{sec:app_details_t_2}
\begin{figure}[t]
    \centering
    \includegraphics[width=\linewidth]{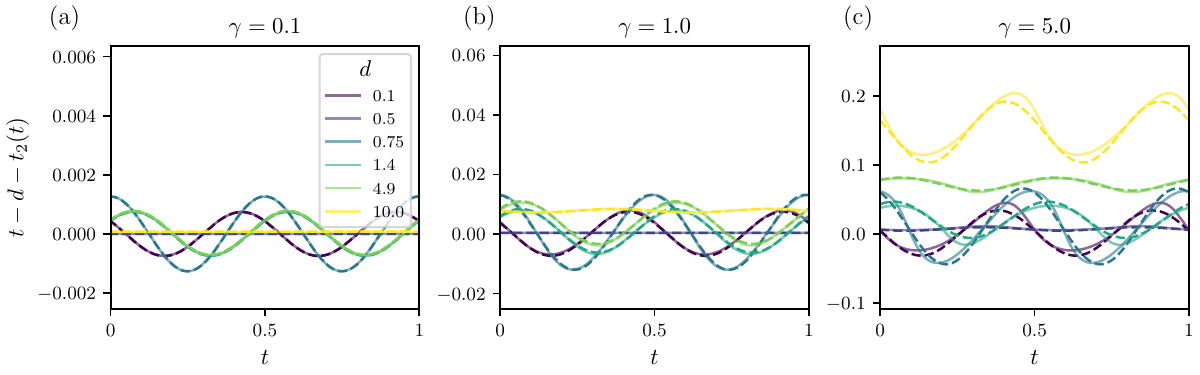}
    \caption{Comparison between the numerical $t_2(t)$ (solid lines) and the asymptotic approximation for $t_2(t)$ (dashed lines), for $\gamma$ equal to (a) 0.1, (b) 1, and (c) 5. } 
    \label{fig:t_2_comparison}
\end{figure}

In this section we use the single-flyer asymptotic solution to approximate $t_2(t)=X_1^{-1}(X_1(t)-d)$. For $\gamma\lesssim1$ we showed that the velocity for a single flyer is well approximated by 
$U_1(t)\approx1+C+R\cos(4\pi t+\varphi)$ where $R$ is given in \eqref{eq:SingleFlyerFirstOrderVelocity} and $C$ is given in \eqref{eq:A0BandC}. Integrating once and using the initial condition $X_1(0)=0$ we obtain 
\begin{align}\label{eq:X1firstorder}
X_1(t)=(1+C)t+\frac{R}{4\pi}[\sin(4\pi t+\varphi)-\sin\varphi].
\end{align}
$t_2(t)$ is given implicitly by $X_1(t_2(t))=X_1(t)-d$, which after substitution of \eqref{eq:X1firstorder} yields
\begin{align}(1+C)t_2(t)+\frac{R}{4\pi}\left[\sin(4\pi t_2(t)+\varphi)-\sin\varphi\right]=(1+C)t+\frac{R}{4\pi}\left[\sin(4\pi t+\varphi)-\sin\varphi\right]-d.
\end{align}
Exploiting the fact that $R(\gamma)\in(0,2/3)$ so that the prefactor in front of the sine is bounded between $0$ and $\sim1/(6\pi)\approx 0.05$, we solve for $t_2(t)$ as an expansion in powers of $R$,
\begin{align}\label{eq:t_2_first_order}
    t_2(t)=t-\frac{d}{1+C}+\frac{R}{2\pi(1+C)}\cos\left(4\pi t+\varphi-\frac{2\pi d}{1+C}\right)\sin\left(\frac{2\pi d}{1+C}\right)+O(R^2).
\end{align}
Thus, to leading order in $R$, $t_2(t)=t-d/(1+C)+O(R)$. In figure~\ref{fig:t_2_comparison} we plot $t-d-t_2(t;\gamma, d)$ for a range of values of $d$ (from 0.1 to 10) and three values of $\gamma$ equal to (a) 0.1, (b) 1, and (c) 5, using both the numerically computed $t_2(t)=X_1^{-1}(X_1(t)-d)$ (solid lines) and the asymptotic approximation given by~\eqref{eq:t_2_first_order} (dashed lines). We plot this quantity instead of $t_2(t)$ directly as it is bounded. We remark that the asymptotic expression approximates the numerical solution extremely well for $\gamma\leq1$ (figures~\ref{fig:t_2_comparison}(a)--(b)) and is still a good approximation for $\gamma=5$ (figure~\ref{fig:t_2_comparison}(c)).

\section{Detailed analysis of separation-enforcing flapping velocities}
In this section we analyse the vertical velocity functions $V_n(t;\gamma,d,\chi)$ for a purely sinusoidal leader using Fourier methods. To this end, we assume $V_n(t;\gamma,d,\chi)$ may be expanded as a Fourier series:
\begin{align}
    V_n(t;\gamma,d,\chi)=\sum_{k\in\mathbb Z} \hat{V}^k_n(\gamma,d,\chi)e^{2\pi i kt}.
\end{align}
For concreteness, we focus on the in-phase functions with $\sigma_n=+1$.
\subsection{Convergence for $n\to\infty$}\label{sec:AppConvergenceInN}
\begin{figure}
    \centering
    \includegraphics[width=\linewidth]{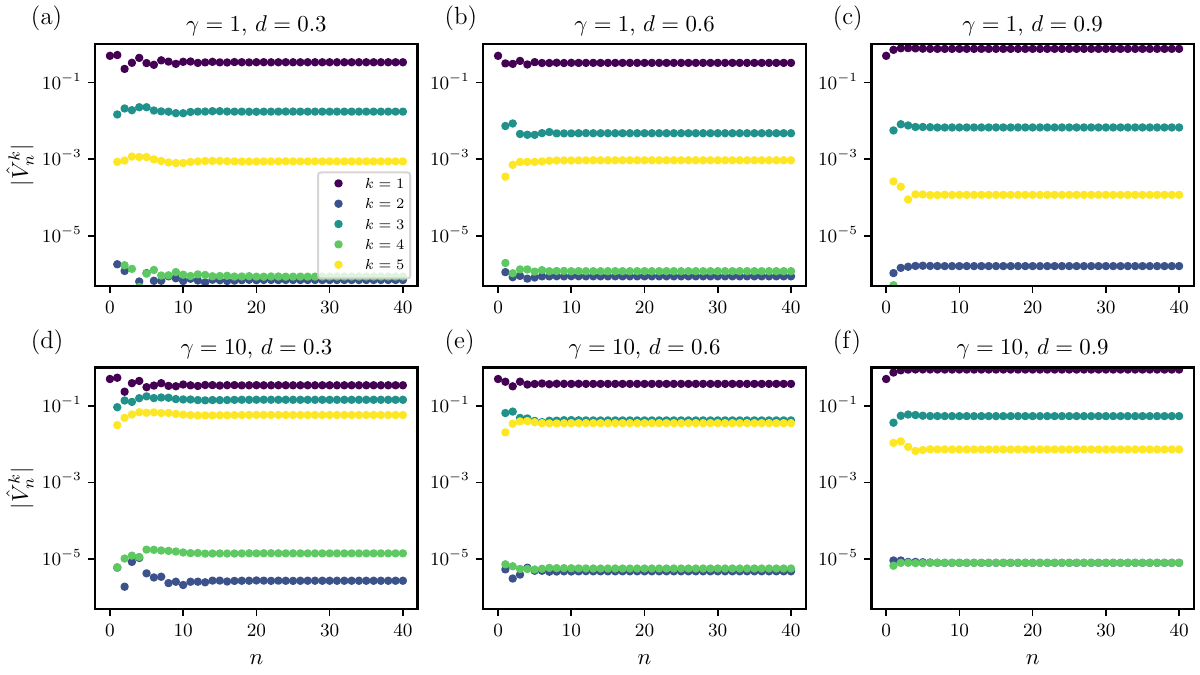}
    \caption{Convergence of the Fourier coefficients for the first five Fourier modes $k=1,2,\dots,5$ as $n\to\infty$ for  $\gamma=1$ (panels (a)--(c)) and $\gamma=10$ (panels (d)--(f)) when flyers  maintain a constant separation distance $d$ (equal to 0.3, 0.6, and 0.9; increasing from left to right).}
    \label{fig:ConvergenceinN}
\end{figure}

We plot in figure~\ref{fig:ConvergenceinN} how the magnitude of $\hat V_n^k(\gamma,d,\chi)$ evolves as $n$ is increased for the same values of $\gamma$ and~$d$ as in figure~\ref{fig:Vn_gamma_d_inphase} in the main text, and with $\chi=1$ fixed. We observe that there is rapid convergence towards the limiting values $\hat V_\infty^k(\gamma,d,\chi)$ as suggested from the approximation for $V_n(t)$ derived in the main text, which points towards an exponentially fast convergence.
\subsection{Even frequencies are skipped}\label{sec:AppFFTinPhase}
\begin{figure}[htpb!]
    \centering
    \includegraphics[width=\linewidth]{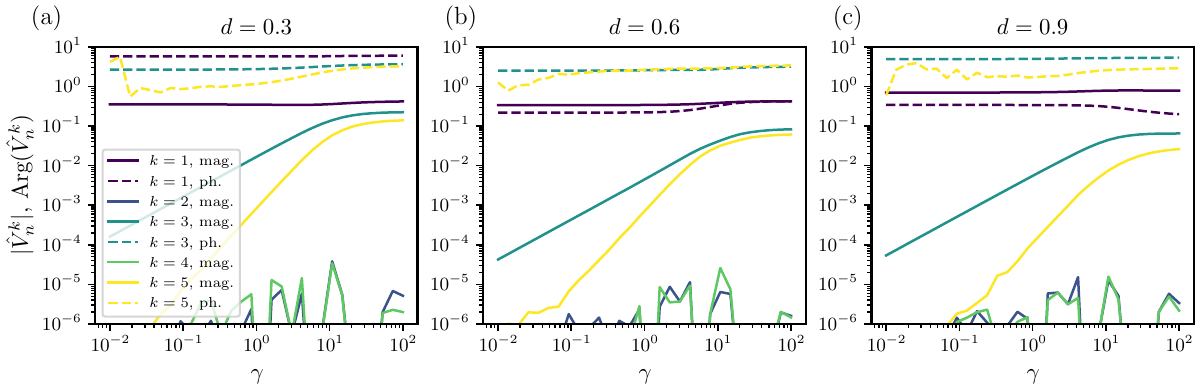}
    \caption{The magnitude $|\hat{V}_n^k|$ for the even coefficients $k=2,4$ remains small across the entire range of $\gamma$, including large values. Odd coefficients grow, as expected from figures~\ref{fig:Vn_gamma_d_inphase} and \ref{fig:Vn_gamma_sweep_out_phase}.}
    \label{fig:higher_order_frequencies}
\end{figure}

We now turn our attention to quantifying the emergence of higher order frequencies in the limiting flapping velocity $V_\infty$ as $\gamma\to\infty$. In figure~\ref{fig:higher_order_frequencies} we plot how the first five Fourier modes, $\hat V_\infty^k( \gamma,d,\chi)$ for $k=1,2,3,4,5$, evolve as $\gamma$ is increased for $d=0.3,0.6,0.9$ and $\chi=1$. The magnitude of $\hat V_\infty^k(\gamma,d,\chi)$ is plotted as a solid line, its phase as a dashed line. We remark that even modes are negligible even for extremely large values of $\gamma$. The first mode retains roughly the same value for all $\gamma$ (in order of magnitude), and we approximate in the main text $\hat V_\infty^1( \gamma,d,\chi)\approx J_\infty/2$, with $J_\infty$ given in \eqref{eq:J_infinity}. However, for the third and fifth modes (and other higher order odd modes) the story is different: their importance is negligible for small $\gamma$ values, but they eventually grow in size and convergence to a finite values as $\gamma\to\infty$, explaining the high-frequency oscillations we observed in figure~\ref{fig:Vn_gamma_d_inphase} of the main text for $\gamma=10$.

We rationalise the appearance of higher order modes by considering a correction to our formula for $V_n$ in \eqref{eq:GeneralFormulaVn} that accounts for larger values of $\gamma$. In particular, expanding $V_n$ as in the main text
\begin{align}
    V_n(t)=V_n^{(0)}+R\,V_n^{(1)}+O(R^2), \qquad R \text{ given in \eqref{eq:SingleFlyerFirstOrderVelocity}},
\end{align}
we find for the large interaction limit ($\chi\to\infty$, the exponentials in \eqref{eq:GeneralFormulaVn} go to unity)
\begin{subequations}
    \begin{align}
    &V_n(t)=\sum_{j=0}^{n-1}\sigma_{n-j}V_1(t_{j+1}(t)),\qquad X_1(t_{j+1}(t))=X_1(t)-jd,\\
    &t_{j+1}(t)=t-\frac{jd}{1+C}+\frac{R}{2\pi(1+C)}\cos\left(4\pi t+\varphi-\frac{2\pi jd}{1+C}\right)\sin\left(\frac{2\pi jd}{1+C}\right)+O(R^2),\\
    &V_n^{(0)}(t)=\sum_{j=0}^{n-1}\sigma_{n-j}\cos\bigl(2\pi (t-j\tilde d)\bigr),\qquad \tilde d=\frac{d}{1+C},\\
    &V_n^{(1)}(t)=\sum_{j=0}^{n-1}\sigma_{n-j}\frac{\upd V_1}{\upd s}\bigg\vert_{s=t_{j+1}^{(0)}(t)}t_{j+1}^{(1)}(t)=-\frac{1}{2\pi(1+C)}\sum_{j=0}^{n-1}\sigma_{n-j}\sin(2\pi j \tilde d)\cos(4\pi t +\varphi -2\pi j\tilde d)\sin(2\pi(t-j\tilde d)).
\end{align}
\end{subequations}
We may convert to complex form to evaluate a    sum to find $V_n^{(1)}(t)$, but this is not necessary to observe that the product of $e^{4\pi it}$ and $e^{2\pi i t}$ terms outputs $e^{2\pi it}$ and $e^{6\pi i t}$ terms, i.e. the first and third fundamental frequencies, with the second fundamental frequency being skipped, as first noted in figure~\ref{fig:higher_order_frequencies}.

\section{Stability analysis of flapping velocities}\label{sec:AppStability}

In a two-flyer system, with the leader flapping sinusoidally with $V_1(t)=\cos(2\pi t)$, the follower must flap with velocity given by
\begin{align}\label{eq:flapping_velocity_follower}
    V_2(t)=\sigma_2 V_1(t)+V_1(X_1^{-1}(X_1(t)-d))e^{-\left(t-X_1^{-1}(X_1(t)-d)\right)/\chi},\qquad \sigma_2=\pm1,
\end{align}
to maintain a constant distance $d$ from the leader. This is precisely~\eqref{eq:RecursiveFormulainTermsoft_2}.
In what follows, we determine which value of $\sigma_2$ (i.e.\ which of the sign choices) is more desirable from the point of view of stability. If the follower is slightly perturbed to a distance $d+\epsilon$ but it keeps flapping according to \eqref{eq:flapping_velocity_follower}, do the perturbations decay so that the follower returns to a distance $d$? 
Alternatively, do the perturbations grow leading to collisions or loss of cohesion due to separation? We take $\sigma_2=\pm 1$ and $\epsilon\ll1$.

\begin{figure}[htpb!]
    \centering
    \includegraphics[width=\linewidth]{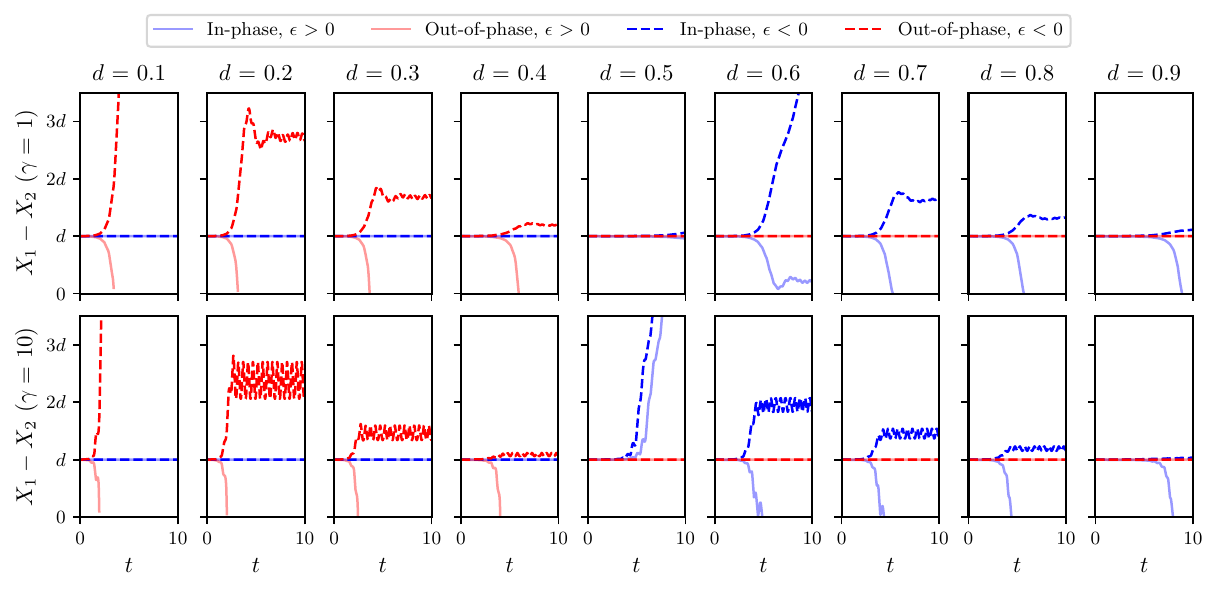}
    \caption{Numerical solutions of \eqref{eq:StabilityEquationFull} for the in-phase flapping ($\sigma_2=+1$, blue) and out-of-phase flapping ($\sigma_2=-1$, red) with separation-distance perturbation $\epsilon>0$ (solid) and $\epsilon<0$ (dashed). Parameter values: $\epsilon=\pm 10^{-4}$, $\chi=1$, $d=0.1,0.2,\dots,0.9$, and $\gamma$ equal to 1 (top row) and 10 (bottom row). The solutions are plotted for $t\in[0,10]$.}
    \label{fig:numerical_stability}
\end{figure}

A force balance on the second flyer yields
\begin{subequations}
\begin{align}
\dot X_2 &=U_2,
\\
\begin{split}
    \frac{1}{\gamma}\dot U_2&=2\left[V_2(t)-V_1(X_1^{-1}(X_2(t)))e^{-\left(t-X_1^{-1}(X_2(t))\right)/\chi}\right]^2-U_2^{3/2}\\&
    =2\left[\sigma_2 V_1(t)+V_1(X_1^{-1}(X_1(t)-d))e^{-\left(t-X_1^{-1}(X_1(t)-d)\right)/\chi}-V_1(X_1^{-1}(X_2(t)))e^{-\left(t-X_1^{-1}(X_2(t))\right)/\chi}\right]^2-U_2^{3/2},\label{eq:StabilityEquationFull}
\end{split}
\end{align}
\end{subequations}
where we substitute for $V_2(t)$ in~\eqref{eq:StabilityEquationFull} using~\eqref{eq:flapping_velocity_follower}.
We now expand the solution for the follower:
\begin{subequations}
    \begin{align}
        X_2(t)=X_2^{(0)}(t)+\epsilon X_2^{(1)}(t)+\dots,\qquad
        U_2(t)=U_2^{(0)}(t)+\epsilon U_2^{(1)}(t)+\dots.
    \end{align}
\end{subequations}
The solution to the leading order problem, is, by construction, $X_2^{(0)}(t)=X_1(t)-d$, $U_2^{(0)}(t)=U_1(t)$, and the first-order problem is 
\begin{align}
    \frac{1}{\gamma}\dot U_2^{(1)}(t)=&-4\sigma_2 e^{-\left(t-X_1^{-1}(X_1(t)-d)\right)/\chi}\left[V_1'(X_1^{-1}(X_1(t)-d))+\frac{1}{\chi}V_1(X_1^{-1}(X_1(t)-d))\right]\cdot\nonumber\\
    &V_1(t)X_2^{(1)}(t)(X_1^{-1})'(X_1(t)-d)-\frac{3}{2}\sqrt{U_1}\,U_2^{(1)}(t),
\end{align}
where the unknowns are $X_2^{(1)}(t)$, $U_2^{(1)}(t)$, and $\dot U_2^{(1)}(t)$. In principle, this constitutes a linear problem that can be solved for a particular choice of $\sigma_2=\pm1$, with initial conditions $X_2^{(1)}(0)=1$ and $U_2^{(1)}(0)=0$, and this is done in figure~\ref{fig:numerical_stability} for $d=0.1, \dots, 0.9$, and $\gamma =1$ (top row) and $10$ (bottom row), as well as $\epsilon =\pm 10^{-4}$. We plot $X_1(t)-X_2(t)$ and observe that the in-phase choice ($\sigma_2=+1$; blue line) is stable for $d=0.1,0.2,0.3, 0.4$, meaning that $X_1(t)-X_2(t)=d$ for all time $t$, and the out-of-phase choice (red line) is stable for $d=0.5,0.6,\dots,0.9$. When unstable, $X_1(t)-X_2(t)$ can either tend to zero at some time, which corresponds to a collision, adjust to a new nearby equilibrium separation that is not equal to $d$, or grow to $X_1(t)-X_2(t)\gg d$.

For small and moderate values of $\gamma\lesssim1$, we have
\begin{subequations}
\begin{align}
U_1(t)&=1+C+R\cos(4\pi t+\varphi),\qquad X_1(t)=(1+C)t+\frac{R}{4\pi}\left[\sin(4\pi t+\varphi)-\sin(\varphi)\right],\\
    X_1^{-1}(x)&=\frac{x}{1+C}-\frac{R}{2\pi(1+C)}\left[\sin\left(\frac{4\pi x}{1+C}+\varphi\right)-\sin\varphi\right]+O(R^2),\\
    X_1^{-1}(X_1(t)-d)&=t-\frac{d}{1+C}+\frac{R}{2\pi(1+C)}\cos\left(4\pi t+\varphi-\frac{2\pi d}{1+C}\right)\sin\left(\frac{2\pi d}{1+C}\right)+O(R^2).
\end{align}
\end{subequations}
with $R,\varphi$ and $C$ (known) functions of $\gamma$ given in \eqref{eq:SingleFlyerFirstOrderVelocity}, and in this regime $R\lesssim1$. We perform a second expansion in powers of $R\ll1$, obtaining to leading-order:
\begin{align}
    \frac{1}{\gamma}\dot U_2^{(1)}=-\sigma\frac{4}{\chi}e^{-d/\chi}\cos(2\pi t)\left[2\pi \chi\sin(2\pi (t-d))+\cos(2\pi(t-d))\right]\,X_2^{(1)}-\frac{3}{2}U_2^{(1)},\qquad \dot X_2^{(1)}=U_2^{(1)}.
\end{align}
This can be written as a single equation for $X_2^{(1)}(t)$, as follows:
\begin{align}
\begin{split}
    &\ddot X_2^{(1)}+\frac{3}{2}\gamma\dot X_2^{(1)}+(p_0+p(t))X_2^{(1)}=0,\\ p_0&=-\sigma \frac{2 \gamma   e^{-d/{\chi }}}{\chi } (2 \pi  \chi  \sin (2 \pi  d)+\cos (2 \pi  d)),\\ p(t)&=-\sigma\frac{2 \gamma    e^{-d/{\chi }}}{\chi } (2 \pi  \chi  \sin (2 \pi  (d-2 t))+\cos (2 \pi  (2 t-d))),
    \end{split}
\end{align}
with the decomposition chosen so that $\int_0^1 p(t)\upd t=0$. Note the fundamental frequency in the forcing is $4\pi$ instead of $2\pi$. 

The system is fundamentally unstable if $p_0<0$ then, in the sense that it is unstable for any damping strength. We see that the sign of 
\begin{align}
    S(d,\chi,\sigma)=-\sigma [\cos(2\pi d)+2\pi \chi\sin(2\pi d)]
\end{align}
is extremely important. We immediately see that if $S<0$ for $\sigma=+1$ then $S>0$ for $\sigma=-1$, so that stability is exchanged between the in-phase and the out-phase flapping velocities. We plot this in figure~\ref{fig:StabilityDiagram}, showing regions where the $\sigma=+1$ flapping velocity is stable to perturbations in blue and regions where the $\sigma=-1$ choice is stable in red. Therefore, the curves in $(d,\chi)$ space where stability changes are
\begin{align}
    \chi_\mathrm{crit}(d)=-\frac{1}{2\pi}\cot(2\pi d).
\end{align}
This line is plotted in black in figure~\ref{fig:StabilityDiagram}, and it is the line of stability exchange. We compare our theoretical results with simulations for different $\gamma$ and $d$ in figure~\ref{fig:numerical_stability}, where we solve the full stability equation \eqref{eq:StabilityEquationFull}.

\section*{Acknowledgments}
\noindent We thank Maria Bruna, Ian Griffiths, Derek Moulton, Alessandro Podo, Leif Ristroph, and Dominic Vella for useful discussions. J.C.V. is funded by St John's College, University of Oxford. C.M. acknowledges funding support provided by a Hooke Research Fellowship at the Mathematical Institute of the University of Oxford. For the purpose of Open Access, the authors will apply a CC BY public copyright license to any Author Accepted Manuscript version arising from this submission.

\bibliographystyle{unsrt}  
\bibliography{biblio.bib}

@article{ramananarivo2016flow,
  title={Flow interactions lead to orderly formations of flapping wings in forward flight},
  author={Ramananarivo, S. and Fang, F. and Oza, A. and Zhang, J. and Ristroph, L.},
  journal={Physical Review Fluids},
  volume={1},
  number={7},
  pages={071201},
  year={2016},
  publisher={APS}
}

@incollection{anderson2008uav,
  title={{UAV} formation control: {Theory} and application},
  author={Anderson, B.~D.~O. and Fidan, B. and Yu, C. and Walle, D.},
  booktitle={Recent advances in learning and control},
  pages={15--33},
  year={2008},
  publisher={Springer}
}

@article{chen2023leader,
  title={Leader-{Follower} {UAV} formation flight control based on feature modelling},
  author={Chen, Y. and Deng, T.},
  journal={Systems Science \& Control Engineering},
  volume={11},
  number={1},
  pages={2268153},
  year={2023},
  publisher={Taylor \& Francis}
}

@inproceedings{ghamry2015formation,
  title={Formation control of multiple quadrotors based on leader-follower method},
  author={Ghamry, K.~A. and Zhang, Y.},
  booktitle={2015 International Conference on Unmanned Aircraft Systems (ICUAS)},
  pages={1037--1042},
  year={2015},
  organization={IEEE}
}

@inproceedings{mercado2013quadrotors,
  title={Quadrotors flight formation control using a leader-follower approach},
  author={Mercado, D.~A. and Castro, R. and Lozano, R.},
  booktitle={2013 European Control Conference (ECC)},
  pages={3858--3863},
  year={2013},
  organization={IEEE}
}

@article{yoon2010cooperative,
  title={Cooperative search and survey using autonomous underwater vehicles ({{AUVs}})},
  author={Yoon, S. and Qiao, C.},
  journal={IEEE Transactions on Parallel and Distributed Systems},
  volume={22},
  number={3},
  pages={364--379},
  year={2010},
  publisher={IEEE}
}

@article{rafifandi2019leader,
  title={Leader--follower formation control of two quadrotor {UAVs}},
  author={Rafifandi, R. and Asri, D.~L. and Ekawati, E. and Budi, E.~M.},
  journal={SN Applied Sciences},
  volume={1},
  number={6},
  pages={539},
  year={2019},
  publisher={Springer}
}

@article{ali2021leader,
  title={A leader-follower formation control of multi-{UAVs} via an adaptive hybrid controller},
  author={Ali, Z.~A. and Israr, A. and Alkhammash, E.~H. and Hadjouni, M.},
  journal={Complexity},
  volume={2021},
  number={1},
  pages={9231636},
  year={2021},
  publisher={Wiley Online Library}
}

@inproceedings{hang2024flow,
  title={Flow sensing and feedback control for maintaining school cohesion in uncoordinated flapping swimmers},
  author={Hang, H. and Heydari, S. and Kanso, E.},
  booktitle={2024 American Control Conference (ACC)},
  pages={3960--3965},
  year={2024},
  organization={IEEE}
}

@article{mavroyiakoumou2025modeling,
  title={Modeling flying formations as flow-mediated matter},
  author={Mavroyiakoumou, C. and Wu, J. and Ristroph, L.},
  journal={arXiv preprint arXiv:2506.14025},
  year={2025}
}

@book{schlichting2016boundary,
  title={Boundary-layer theory},
  author={Schlichting, H. and Gersten, K.},
  year={2016},
  publisher={springer}
}

@article{han2025tailoring,
  title={Tailoring formations of self-organising hydrofoil schools towards high-efficiency},
  author={Han, T. and Mivehchi, A. and Sarraf, S. Seyedmirzaei and Moored, K.~W.},
    volume={1012}, 
    DOI={10.1017/jfm.2025.10239}, journal={Journal of Fluid Mechanics}, 
    year={2025}, 
    pages={A26}
}

@article{andersen2017wake,
  title={Wake structure and thrust generation of a flapping foil in two-dimensional flow},
  author={Andersen, A. and Bohr, T. and Schnipper, T. and Walther, J.~H.},
  journal={Journal of Fluid Mechanics},
  volume={812},
  pages={R4},
  year={2017},
  publisher={Cambridge University Press}
}

@article{smits2019undulatory,
  title={Undulatory and oscillatory swimming},
  author={Smits, A.~J.},
  journal={Journal of Fluid Mechanics},
  volume={874},
  pages={P1},
  year={2019},
  publisher={Cambridge University Press}
}

@article{rival2011recovery,
  title={Recovery of energy from leading-and trailing-edge vortices in tandem-airfoil configurations},
  author={Rival, D. and Hass, G. and Tropea, C.},
  journal={Journal of Aircraft},
  volume={48},
  number={1},
  pages={203--211},
  year={2011}
}

@article{boschitsch2014propulsive,
  title={Propulsive performance of unsteady tandem hydrofoils in an in-line configuration},
  author={Boschitsch, B.~M. and Dewey, P.~A. and Smits, A.~J.},
  journal={Physics of Fluids},
  volume={26},
  number={5},
  year={2014},
  publisher={AIP Publishing}
}

@article{major1978three,
  title={The three-dimensional structure of airborne bird flocks},
  author={Major, P.~F. and Dill, L.~M.},
  journal={Behavioral Ecology and Sociobiology},
  volume={4},
  number={2},
  pages={111--122},
  year={1978},
  publisher={Springer}
}

@article{gould1974vee,
  title={The vee formation of Canada geese},
  author={Gould, L.~L. and Heppner, F.},
  journal={The Auk},
  pages={494--506},
  year={1974},
  publisher={JSTOR}
}

@article{cade2020predator,
  title={Predator-informed looming stimulus experiments reveal how large filter feeding whales capture highly maneuverable forage fish},
  author={Cade, D. E. and Carey, N. and Domenici, P. and Potvin, J. and Goldbogen, J. A.},
  journal={Proceedings of the National Academy of Sciences},
  volume={117},
  number={1},
  pages={472--478},
  year={2020},
  publisher={National Academy of Sciences}
}

@article{goldbogen2017baleen,
  title={How baleen whales feed: the biomechanics of engulfment and filtration},
  author={Goldbogen, J.~A. and Cade, D.~E. and Calambokidis, J. and Friedlaender, A.~S. and Potvin, J. and Segre, P.~S. and Werth, A.~J.},
  journal={Annual review of marine science},
  volume={9},
  number={1},
  pages={367--386},
  year={2017},
  publisher={Annual Reviews}
}

@article{fang2025flowinteractionsforwardflight,
  title={Flow interactions and forward flight dynamics of tandem flapping wings},
  author={Fang, F. and Mavroyiakoumou, C. and Ristroph, L. and Shelley, M.~J.},
  journal={arXiv preprint arXiv:2505.13149},
  year={2025},
doi={
https://doi.org/10.48550/arXiv.2505.13149}
}

@article{cavagna2014bird,
  title={Bird flocks as condensed matter},
  author={Cavagna, A. and Giardina, I.},
  journal={Annu. Rev. Condens. Matter Phys.},
  volume={5},
  number={1},
  pages={183--207},
  year={2014},
  publisher={Annual Reviews}
}

@article{hemelrijk2012schools,
  title={Schools of fish and flocks of birds: their shape and internal structure by self-organization},
  author={Hemelrijk, C.~K. and Hildenbrandt, H.},
  journal={Interface focus},
  volume={2},
  number={6},
  pages={726--737},
  year={2012},
  publisher={The Royal Society}
}

@article{heydari2021school,
  title={School cohesion, speed and efficiency are modulated by the swimmers flapping motion},
  author={Heydari, S. and Kanso, E.},
  journal={Journal of Fluid Mechanics},
  volume={922},
  pages={A27},
  year={2021},
  publisher={Cambridge University Press}
}

@article{sumpter2006principles,
  title={The principles of collective animal behaviour},
  author={Sumpter, D.~J.~T.},
  journal={Philosophical transactions of the Royal Society B: Biological Sciences},
  volume={361},
  number={1465},
  pages={5--22},
  year={2006},
  publisher={The Royal Society London},
  doi={https://doi.org/10.1098/rstb.2005.1733}
}

@article{floryan2017scaling,
  title={Scaling the propulsive performance of heaving and pitching foils},
  author={Floryan, D. and Van Buren, T. and Rowley, C.~W. and Smits, A.~J.},
  journal={Journal of Fluid Mechanics},
  volume={822},
  pages={386--397},
  year={2017},
  publisher={Cambridge University Press}
}

@article{triantafyllou1993optimal,
  title={Optimal thrust development in oscillating foils with application to fish propulsion},
  author={Triantafyllou, G.~S. and Triantafyllou, M.~S. and Grosenbaugh, M.~A.},
  journal={Journal of Fluids and Structures},
  volume={7},
  number={2},
  pages={205--224},
  year={1993},
  publisher={Elsevier}
}

@article{oza2019lattices,
  title={Lattices of hydrodynamically interacting flapping swimmers},
  author={Oza, A.~U. and Ristroph, L. and Shelley, M.~J.},
  journal={Physical Review X},
  volume={9},
  number={4},
  pages={041024},
  year={2019},
  publisher={APS}
}

@article{vandenberghe2004symmetry,
  title={Symmetry breaking leads to forward flapping flight},
  author={Vandenberghe, N. and Zhang, J. and Childress, S.},
  journal={Journal of Fluid Mechanics},
  volume={506},
  pages={147--155},
  year={2004},
  publisher={Cambridge University Press}
}

@article{vandenberghe2006unidirectional,
  title={On unidirectional flight of a free flapping wing},
  author={Vandenberghe, N. and Childress, S. and Zhang, J.},
  journal={Physics of Fluids},
  volume={18},
  number={1},
  pages={014102},
  year={2006},
  publisher={American Institute of Physics}
}

@book{tritton2012physical,
  title={Physical fluid dynamics},
  author={Tritton, D.~J.},
  year={2012},
  publisher={Springer Science \& Business Media}
}

@article{becker2015hydrodynamic,
  title={Hydrodynamic schooling of flapping swimmers},
  author={Becker, A.~D. and Masoud, H. and Newbolt, J.~W. and Shelley, M. and Ristroph, L.},
  journal={Nature Communications},
  volume={6},
  number={1},
  pages={1--8},
  year={2015},
  publisher={Nature Publishing Group}
}

@article{newbolt2019flow,
  title={Flow interactions between uncoordinated flapping swimmers give rise to group cohesion},
  author={Newbolt, J.~W. and Zhang, J. and Ristroph, L.},
  journal={Proceedings of the National Academy of Sciences},
  volume={116},
  number={7},
  pages={2419--2424},
  year={2019},
  publisher={National Acad Sciences}
}

@article{spagnolie2010surprising,
  title={Surprising behaviors in flapping locomotion with passive pitching},
  author={Spagnolie, S.~E. and Moret, L. and Shelley, M.~J. and Zhang, J.},
  journal={Physics of Fluids},
  volume={22},
  number={4},
  pages={041903},
  year={2010},
  publisher={American Institute of Physics}
}

@article{alben2005coherent,
  title={Coherent locomotion as an attracting state for a free flapping body},
  author={Alben, S. and Shelley, M.},
  journal={Proceedings of the National Academy of Sciences},
  volume={102},
  number={32},
  pages={11163--11166},
  year={2005},
  publisher={National Acad Sciences}
}

@book{lighthill1975mathematical,
  title={Mathematical biofluiddynamics},
  author={Lighthill, Sir J.},
  year={1975},
  publisher={SIAM}
}

@article{dai2018stable,
  title={Stable formations of self-propelled fish-like swimmers induced by hydrodynamic interactions},
  author={Dai, L. and He, G. and Zhang, X. and Zhang, X.},
  journal={Journal of The Royal Society Interface},
  volume={15},
  number={147},
  pages={20180490},
  year={2018},
  publisher={The Royal Society}
}

@article{peng2018hydrodynamic,
  title={Hydrodynamic schooling of multiple self-propelled flapping plates},
  author={Peng, Z.-R. and Huang, H. and Lu, X.-Y.},
  journal={Journal of Fluid Mechanics},
  volume={853},
  pages={587--600},
  year={2018},
  publisher={Cambridge University Press}
}

@article{nitsche2025stability,
  title={On the stability of an in-line formation of hydrodynamically interacting flapping plates},
  author={Nitsche, M. and Oza, A.~U. and Siegel, M.},
  journal={Journal of Fluid Mechanics},
  volume={1013},
  pages={A14},
  year={2025},
  publisher={Cambridge University Press}
}

@article{saadat2021hydrodynamic,
  title={Hydrodynamic advantages of in-line schooling},
  author={Saadat, M. and Berlinger, F. and Sheshmani, A. and Nagpal, R. and Lauder, G.~V. and Haj-Hariri, H.},
  journal={Bioinspiration \& Biomimetics},
  volume={16},
  number={4},
  pages={046002},
  year={2021},
  publisher={IOP Publishing}
}

@misc{ioc_drone_show_2021,
  author       = {{International Olympic Committee}},
  title        = {{Spectacular Intel Drone Light Show Helps Bring Tokyo 2020 to Life}},
  year         = {2021},
  month        = jul,
  howpublished = {\url{https://www.olympics.com/ioc/news/spectacular-intel-drone-light-show-helps-bring-tokyo-2020-to-life-1}},
  note         = {Accessed: 22 December 2025}
}

@misc{ars_drone100_2016,
  author       = {{ARS Electronica Futurelab}},
  title        = {Drone 100: A World Record with 100 Points},
  year         = {2016},
  month        = jan,
  howpublished = {\url{https://ars.electronica.art/aeblog/en/2016/01/12/drone100/}},
  note         = {Accessed: 22 December 2025}
}

@article{zhihao2020virtual,
  title={Virtual target guidance-based distributed model predictive control for formation control of multiple {UAVs}},
  author={Cai, Z. and Wang, L. and Zhao, J. and Wu, K. and Wang, Y.},
  journal={Chinese Journal of Aeronautics},
  volume={33},
  number={3},
  pages={1037--1056},
  year={2020},
  publisher={Elsevier}
}

@inproceedings{lanteigne2017design,
  title={Design of a drone lead-follow control system},
  author={Lanteigne, A. and Kibru, E. and Azam, S. and Al Shammary, S.},
  booktitle={2017 Systems and Information Engineering Design Symposium (SIEDS)},
  pages={162--167},
  year={2017},
  organization={IEEE}
}

@inproceedings{hagen2019class,
  title={Class-based {ODE} solvers and event detection in {SciPy}},
  author={Hagen, D. and Mayorov, N.},
  booktitle={Python Science Conf.},
  year={2019}
}

@article{zhu2014flow,
  title={Flow-mediated interactions between two self-propelled flapping filaments in tandem configuration},
  author={Zhu, X. and He, G. and Zhang, X.},
  journal={Physical Review Letters},
  volume={113},
  number={23},
  pages={238105},
  year={2014},
  publisher={APS}
}

@article{newbolt2024flow,
  title={Flow interactions lead to self-organized flight formations disrupted by self-amplifying waves},
  author={Newbolt, J.~W. and Lewis, N. and Bleu, M. and Wu, J. and Mavroyiakoumou, C. and Ramananarivo, S. and Ristroph, L.},
  journal={Nature Communications},
  volume={15},
  number={3462},
  pages={1--12},
  year={2024},
  publisher={Springer Nature},
    doi={https://doi.org/10.1038/s41467-024-47525-9}
}

@article{chico-vazquez_mathematical_2025,
    title = {A mathematical model for optimal breakaways in cycling: balancing energy expenditure and crash risk},
    volume = {12},
    shorttitle = {A mathematical model for optimal breakaways in cycling},
    url = {https://royalsocietypublishing.org/doi/10.1098/rsos.250972},
    doi = {10.1098/rsos.250972},
    number = {11},
    urldate = {2025-11-14},
    journal = {Royal Society Open Science},
    author = {Chico-Vázquez, J. and Griffiths, I. M.},
    month = nov,
    year = {2025},
    pages = {250972},
}

@book{emde_senses_2012,
    title = {The {Senses} of {Fish}: {Adaptations} for the {Reception} of {Natural} {Stimuli}},
    isbn = {978-94-007-1060-3},
    shorttitle = {The {Senses} of {Fish}},
    language = {en},
    publisher = {Springer Science \& Business Media},
    author = {von der Emde, G. and Mogdans, J. and Kapoor, B. G.},
    month = dec,
    year = {2012},
}

@article{chico-vazquez_uncovering_2025,
    title = {Uncovering flow and deformation regimes in the coupled fluid–solid vestibular system},
    volume = {1022},
    issn = {0022-1120, 1469-7645},
    url = {https://www.cambridge.org/core/journals/journal-of-fluid-mechanics/article/uncovering-flow-and-deformation-regimes-in-the-coupled-fluidsolid-vestibular-system/7B7109ED170BCE3DF37004BE6C772979},
    doi = {10.1017/jfm.2025.10783},
    language = {en},
    urldate = {2025-11-14},
    journal = {Journal of Fluid Mechanics},
    author = {Chico-Vázquez, J. and Moulton, D. E. and Vella, D.},
    month = nov,
    year = {2025},
    pages = {A40},
}

@book{stein_shakarchi_fourier,
  author    = {Stein, E.~M. and Shakarchi, R.},
  title     = {Fourier Analysis: An Introduction},
  series    = {Princeton Lectures in Analysis},
  volume    = {1},
  publisher = {Princeton University Press},
  year      = {2003},
  isbn      = {978-0691113845}
}

\end{document}